\documentclass[%
superscriptaddress,
preprint,
 amsmath,amssymb,
 aps,
prb,
showkeys
]{revtex4-2}

\usepackage{graphicx}
\usepackage{dcolumn}
\usepackage{bm}
\usepackage{natbib}
\usepackage{afterpage}
\usepackage{color}
\usepackage{ulem}
\usepackage{comment}
\usepackage{ulem}
\newcommand{\pt}{\partial}
\newcommand{\red}[1]{\textcolor{black}{#1}}
\newcommand{\blue}[1]{\textcolor{black}{#1}}

\begin{document}


\title{Synchronized molecular dynamics method for thin-layer flows of complex fluids}

\author{Shugo Yasuda}%
 \email{yasuda@gsis.u-hyogo.ac.jp}
\affiliation{Department of Data and Computational Science, University of Hyogo, Kobe 650-0047, Japan}

\author{Kotaro Oda}
\affiliation{Department of Data and Computational Science, University of Hyogo, Kobe 650-0047, Japan}
\affiliation{Production Technology R \& D Cross-DX Center, R \& D Headquarters, Daicel Corporation, Himeji 671-1283, Japan}

\author{Fumito Muragaki}
\affiliation{Department of Data and Computational Science, University of Hyogo, Kobe 650-0047, Japan}

\author{Yuta Taketa}
\affiliation{Department of Data and Computational Science, University of Hyogo, Kobe 650-0047, Japan}

\author{Masashi Iwayama}
\email{dc11794@daicelgroup.com}
\affiliation{Production Technology R \& D Cross-DX Center, R \& D Headquarters, Daicel Corporation, Himeji 671-1283, Japan}

\author{Tomohide Ina}
\affiliation{Production Technology R \& D Cross-DX Center, R \& D Headquarters, Daicel Corporation, Himeji 671-1283, Japan}


\date{\today}

\begin{abstract}
We propose a multiscale computational method for thin-layer flows of complex fluids, termed the synchronized molecular dynamics (SMD) method, which directly couples local molecular dynamics (MD) simulations with a macroscopic lubrication description. In thin layers, the flow can be decomposed into cross-sectional dynamics that are strongly influenced by interfacial effects, and streamwise transport along the channel. The SMD method exploits this separation of scales by sparsely distributing local MD cells along the channel and synchronizing them through macroscopic conservation laws.

In this framework, the macroscopic continuity equation is enforced by iteratively updating the \red{numerical driving} forces applied to each MD cell, thereby allowing the cross-sectional velocity profiles and the streamwise pressure distribution to be obtained without prescribing constitutive relations or boundary conditions. The method is validated for pressure-driven and wall-driven flows of Lennard--Jones fluids in a wedge-shaped channel, demonstrating excellent agreement with a modified Reynolds equation that accounts for boundary slip.

The SMD method is further applied to polymeric lubrication flows modeled by the Kremer--Grest chain model. At large pressure differences, the present approach naturally captures pronounced shear-thinning behavior coupled with microscopic polymer conformation dynamics. The results demonstrate that the SMD method provides an efficient and physically consistent framework for the multiscale simulation of complex fluid thin-layer flows.
\end{abstract}

\maketitle
\newpage


\section{Introduction}\label{sec:intro}
Flows of soft matter, such as polymers, colloids, and liquid crystals, are ubiquitous in material processing in chemical engineering. In particular, thin-layer flows, including film and lubrication flows, are widely encountered. These fluids exhibit not only complex thermorheological properties but also rich interfacial behaviors. 
\blue{As the physical scale decreases, increased surface-to-volume ratios make surface forces dominant over volume forces. Therefore, the fluid behavior near walls is governed by wall effects. One such effect is wall slip, which is commonly observed in polymer melts. As reviewed by Hatzikiriakos \cite{hatzikiriakos2012wall}, polymer melts exhibit slip above a critical wall shear stress. A secondary critical stress then drives a transition from weak to strong slip, resulting from the disentanglement of bulk chains at the wall. B{\"a}umchen and Jacobs \cite{baumchen2010slip} provided a comprehensive review of fundamental models and experimental findings regarding polymer slip, summarizing the concepts of fluid slippage at the molecular scale.
These slip effects significantly influence the velocity and pressure distributions in the bulk region and should therefore be treated explicitly, particularly in microchannel flows \cite{chien2005study}. However, in conventional computational fluid dynamics, the models and boundary conditions governing these effects are often not well established.} 
Consequently, predicting such flows remains a significant challenge in chemical engineering, soft-matter physics, and computational science. In this paper, we develop a multiscale computational approach that addresses this challenge by directly coupling molecular dynamics with a macroscopic thin-layer flow description.

In molecular dynamics (MD) simulations, constitutive relations and macroscopic boundary conditions are not prescribed explicitly; instead, molecular interactions are resolved directly. \blue{Therefore, MD simulations have been widely employed to investigate fluid transport, such as the flow of gases and water in confined nanochannels~\cite{Barisik2014,Zhou2024}. Furthermore, MD simulations have been applied to the flow dynamics of soft matter~\cite{CHEN2020145284,qiu2024variations,priezjev2004molecular,Yang2013,Mackay2014,cho2017molecular,yildirim2024molecular}.} As a result, MD simulations can capture the thermorheological properties and interfacial behaviors of complex fluids at the microscopic level. However, despite these advantages, direct MD simulations are limited in terms of the spatial and temporal scales they can access, rendering their application to macroscopic flows computationally impractical.

Nevertheless, when the thickness of a thin layer becomes sufficiently small---approaching microscopic scales (e.g., on the order of micrometers)---the flow can be effectively decomposed into local cross-sectional dynamics and macroscopic transport along the channel. The cross-sectional flows, which are strongly influenced by interfacial phenomena at the channel boundaries, can be resolved by MD simulations, while the streamwise transport is described by a simplified macroscopic transport equation. This separation of scales enables computational approaches that retain essential microscopic physics while efficiently describing macroscopic flow behavior.
In the present multiscale framework, instead of performing a single large-scale MD simulation over the entire domain, we sparsely distribute multiple local MD cells along the channel to resolve the local cross-sectional flows and couple them through a macroscopic thin-layer flow description.

Multiscale simulations of complex fluid flows have been extensively developed to address the difficulty of prescribing the constitutive relations.
A seminal approach is the CONNFFESSIT method for polymeric liquids introduced by Laso, Ãttinger, and coworkers\cite{Laso1993,Feigl1995,Laso1997}, in which the local stress in a continuum solver is evaluated directly from microscopic simulations rather than from prescribed constitutive models.

Related strategies have been incorporated into broader multiscale methodologies.
In particular, the heterogeneous multiscale method (HMM) proposed by 
E and Engquist~\cite{Weinan2003} and HMM has been applied to a wide range of flow systems, including polymeric liquids~\cite{Ren2005,E2007,De2006,Yasuda2008,Yasuda2010}.
The equation-free approach developed by Kevrekidis et al.~\cite{Kevrekidis2003} is based on a similar philosophy and has also been successfully applied to diverse multiscale problems~\cite{Cisternas2004,Erban2006}.

With respect to polymeric fluids, Murashima and Taniguchi developed a Lagrangian scheme to capture the advection of the memory of the polymer conformation~\cite{murashima_taniguchi_2010} and successfully applied it to history-dependent flows of polymeric liquids with high Weissenberg numbers~\cite{murashima2011multiscale,murashima_taniguchi_2012,SATO201734,sato2019multiscale,Morri2021}.
Moreno and Ellero proposed a fully Lagrangian heterogeneous multiscale method (LHMM) for simulating complex fluids whose microscopic structures evolve over wide spatial and temporal scales, including polymer solutions and multiphase systems~\cite{moreno2023generalized,patino2025lagrangian}.
We have also developed a multiscale method for polymeric fluids in high-Prandtl-number lubrication flows, where local MD simulations that resolve thermorheological properties are synchronized via macroscopic heat and momentum transport, forming a synchronized molecular dynamics (SMD) framework~\cite{Yasuda2014,2025_IOP_OY}.
Unlike conventional approaches based on constitutive modeling, this method directly captures thermorheological coupling arising from microscopic dynamics.

Despite these developments, multiscale simulations for thin-layer flows of complex fluids, particularly those involving strong interfacial effects, remain limited.
In such systems, the flow exhibits strong coupling between microscopic interfacial dynamics and macroscopic transport along the channel, making it difficult to describe the system solely in terms of macroscopic variables.
\red{To address this issue, we extend our previous SMD framework by introducing cross-sectional MD cells that contain the entire gap, including the fluid--wall interfaces and the atomistic walls, at sparsely distributed streamwise locations.
Unlike our previous SMD simulations, where local MD cells were assigned to the bulk fluid region excluding the confining walls, the present configuration directly resolves interfacial molecular dynamics, wall slip, and cross-sectional stress responses within each local MD cell.}

In the context of thin-layer flows, a multiscale method termed the internal-flow multiscale method (IMM) was proposed by Borg et al.~\cite{Borg2013}, and it enables the treatment of internal flows of simple liquid or rarefied gases in high-aspect-ratio geometries by resolving molecular-scale effects, including slip flow and temperature jump, at solid boundaries
~\cite{Borg2013,PATRONIS2014532,Borg_Lockerby_Reese_2015,john2018simulation}.
Our approach shares technical similarities with IMM, while extending it to treat soft-matter flows that exhibit rich interfacial phenomena and non-Newtonian rheology, thereby introducing additional complexities beyond those encountered in simple liquid or rarefied gas flows.
\red{In particular, the local MD cells in the present SMD method resolve the cross-sectional distribution of polymer conformations and its near-wall variation, together with the local velocity and stress fields. 
The synchronization scheme also incorporates the streamwise variation of the normal stress difference into the macroscopic lubrication constraint, thereby coupling microscopic
conformation-dependent stress responses with the macroscopic pressure distribution.}

In this paper, we develop an SMD method for thin-layer flows of soft matter that directly couples local MD simulations with a macroscopic thin-film flow description.
The structure of this article is described below. After the introduction, the SMD scheme and its implementation are described. The subsequent sections examine the validity and efficiency of the SMD method. Pressure-driven and wall-driven flows of Lennard--Jones fluids in a wedge-shaped channel are analyzed over a range of fluid densities and fluid--wall interaction strengths, and the results are compared with those of a modified Reynolds equation that accounts for slip at the boundaries. The SMD method is then applied to pressure-driven flows of a polymeric liquid with the same geometry, demonstrating the non-Newtonian behavior and interfacial dynamics of the polymer conformation. Finally, concluding remarks are presented.

\section{Synchronized molecular dynamics method}

\begin{figure}[t]
\centering
\includegraphics[width=0.8\textwidth]{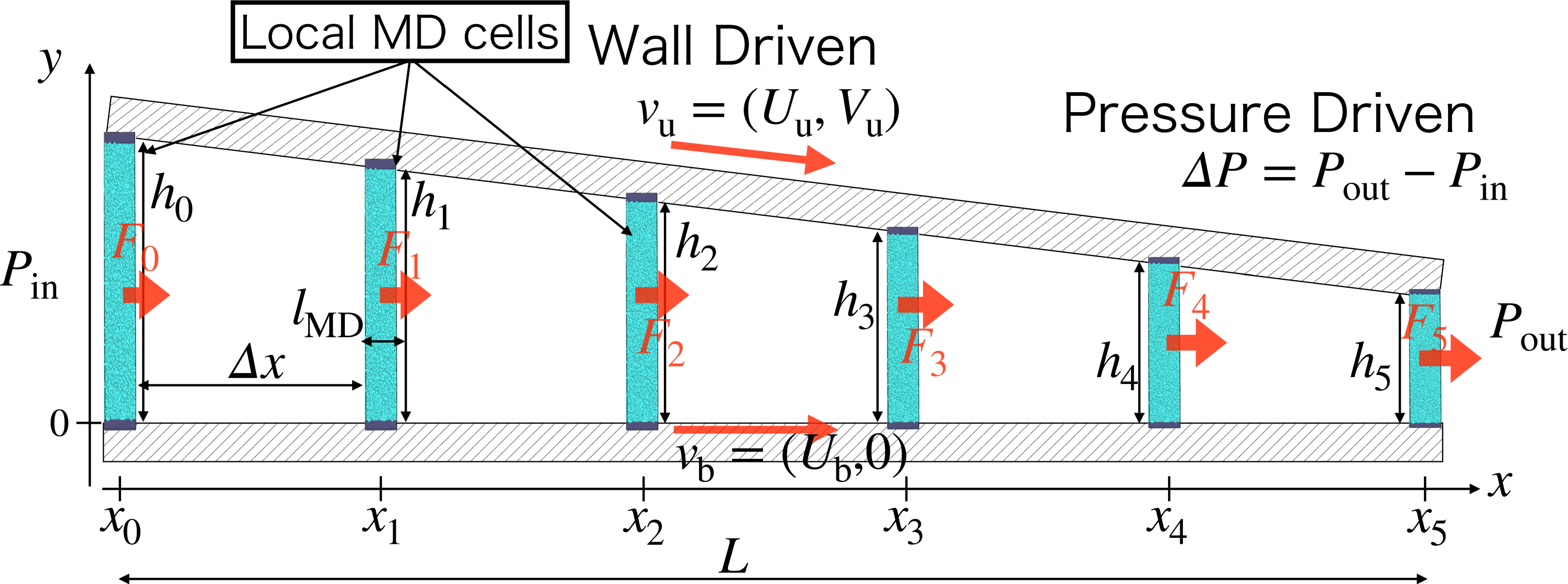}
\caption{Geometry of the problem. Laminar flow in a narrow gap is considered, where local MD cells are arranged at uniform intervals of $\varDelta x$ along the channel.
MD cells are rectangular boxes with a base of $l_\mathrm{MD}\times l_\mathrm{MD}$.
\red{Only these rectangular local MD cells are explicitly simulated by MD; the intervening portions of the channel are not directly resolved at the molecular level but are coupled through the macroscopic lubrication description.}
In each MD cell, the wall and fluid are modeled with Lennard-Jones molecules, and the fluid molecules are driven by \red{numerical driving} forces $F_i$ ($i=0,\cdots,I$) that are adjusted to satisfy the macroscopic continuum equation throughout the channel.}\label{fig:geom}
\end{figure}

\subsection{Macroscopic description}
We consider laminar flow in a narrow gap (as depicted in Fig.~\ref{fig:geom}) under the following assumptions:
\begin{itemize}
\item the height of the gap is much smaller than the channel length, i.e., the aspect ratio $\red{\chi}=H_\mathrm{c}/L\ll 1$,
\item the flow is laminar with a low Reynolds number, i.e., $Re=U_\mathrm{c} H_\mathrm{c}/\nu_\mathrm{c} \ll 1$,
\item the fluid is incompressible, i.e., $\nabla\cdot \bm{v}=0$,
\end{itemize}
where 
$H_\mathrm{c}$ is the characteristic height of the gap \red{satisfying $\min_x h(x)\le H_\mathrm{c}\le \max_xh(x)$} (see also Fig.~\ref{fig:geom}),
$\bm{v}$ is the flow velocity, 
$U_\mathrm{c}$ is the characteristic flow velocity along the channel, 
and $\nu_\mathrm{c}$ is the kinematic viscosity.
\red{
Under these assumptions, variations in the cross-sectional direction dominate those in the streamwise direction because of the small aspect ratio.}
For two-dimensional laminar flows with $\bm{v}=(u,v,0)$,
\red{neglecting the higher-order terms of $O(\chi^2)$ and $O(\chi Re)$ in the momentum equations yields the following leading-order lubrication equations}  (see Appendix〜A):
\begin{gather}
\frac{\pt u}{\pt x}+\frac{\pt v}{\pt y}=0\label{eq_cont},\\
\rho \frac{\partial u}{\partial t}
=
\frac{\pt \sigma_{xx}}{\pt x}
+\frac{\pt \sigma_{yx}}{\pt y},\label{eq_moment}\\
\frac{\pt \sigma_{yy}}{\pt y}=0.\label{eq_pyy}
\end{gather}
Here, $\rho$ is the fluid density, and $\sigma_{\alpha\beta}$ is the stress tensor.
Hereafter, the subscripts $\alpha$ and $\beta$ represent Cartesian coordinates, i.e., $\{\alpha,\beta\}=\{x,y\}$.
\red{We note that, in the present formulation, the incompressibility assumption is imposed
at the macroscopic level in the kinematic sense, i.e., $\nabla \cdot \bm{v}=0$.
The validity of this approximation requires that the characteristic flow velocity $U_\mathrm{c}$, in particular the streamwise velocity in the lubrication approximation, be sufficiently small compared with the speed of sound $c_\mathrm{s}$, i.e., $Ma=U_\mathrm{c}/c_\mathrm{s} \ll 1$. 
This low-Mach-number condition is expected to be satisfied in many thin-layer flows of polymeric liquids and other soft-matter fluids encountered in engineering lubrication processes.
On the other hand, flows involving strong compressibility, density-wave propagation, or shock-wave structures are outside the scope of the present SMD formulation.
We also restrict our attention to the case in which the wall temperature is spatially uniform and temporally constant, and the streamwise temperature inhomogeneity caused by local viscous heating remains sufficiently small.
Therefore, macroscopic energy transport in the streamwise direction is not considered in the present formulation.
Consequently, the present SMD method does not resolve macroscopic streamwise variations in density or temperature, although molecular-scale density inhomogeneities and small temperature variations in the cross-sectional direction may exist within each local MD cell.
}

As the boundary condition of Eq.~(\ref{eq_moment}) at the upper and lower walls (i.e., $y=h_\mathrm{u}(x)$, $y=h_\mathrm{b}(x)$, respectively), we consider the nonflux condition in the normal direction on the wall surface, i.e.,
\begin{equation}\label{eq_nonflux}
\begin{split}
(u(x,h_\mathrm{u})-U_\mathrm{u}(x),v(x,h_\mathrm{u})-V_\mathrm{u}(x))\cdot
\left(
\begin{array}{c}
-h_\mathrm{u}'(x)\\
1
\end{array}
\right)=0,\\
(u(x,h_\mathrm{b})-U_\mathrm{b}(x),v(x,h_\mathrm{b})-V_\mathrm{b}(x))\cdot
\left(
\begin{array}{c}
h_\mathrm{b}'(x)\\
-1
\end{array}
\right)=0,
\end{split}
\end{equation}
where $(U_\mathrm{u},V_\mathrm{u})$ and $(U_\mathrm{b},V_\mathrm{b})$ are the velocities of the upper and lower walls, respectively, and the prime ${}'$ represents the derivative with respect to $x$.

By integrating Eq.~(\ref{eq_cont}) with respect to $y\in [h_\mathrm{b}(x),h_\mathrm{u}(x)]$ at each position $x$ and using the nonflux condition (\ref{eq_nonflux}), we obtain the following continuity equation:
\begin{align}\label{eq_cont2}
&\int_{h_\mathrm{b}(x)}^{h_\mathrm{u}(x)}\frac{\pt u}{\pt x}dy
=-\int_{h_\mathrm{b}(x)}^{h_\mathrm{u}(x)}\frac{\pt v}{\pt y}dy,\nonumber \\
&\frac{\pt}{\pt x}\left[\int_{h_\mathrm{b}(x)}^{h_\mathrm{u}(x)}u(x,y) dy\right]
-u(x,h_\mathrm{u})h'_u(x)+u(x,h_\mathrm{b})h'_b(x)=V_\mathrm{b}(x)-V_\mathrm{u}(x),\nonumber \\
&M'(x)=U_\mathrm{u}(x)h'_\mathrm{u}(x)-U_\mathrm{b}(x)h'_\mathrm{b}(x)
+V_\mathrm{b}(x)-V_\mathrm{u}(x),
\end{align}
 where
\begin{equation}\label{eq_flux}
M(x)=\int_{h_\mathrm{b}(x)}^{h_\mathrm{u}(x)}u(x,y)dy.
\end{equation}

We note that Eq.~(\ref{eq_cont2}) remains valid even for the slip boundary condition, as the derivation of Eq.~(\ref{eq_cont2}) does not consider the nonslip boundary condition but only the nonflux condition in the normal direction on the wall surface.
We also note that the right-hand side of Eq.~(\ref{eq_cont2}) vanishes when the channel wall moves only in the tangential direction along the wall (i.e., $V_\mathrm{u,b}/U_\mathrm{u,b}=h'_\mathrm{u,b}(x)$).

Below, we consider only the flat-bottom plate, i.e., $h_\mathrm{b}(x)=0$ and $V_\mathrm{b}=0$, and write the channel height at position $x$ as $h(x)$.

For the Newtonian fluid, which has a constant viscosity $\mu$ (i.e., $\sigma_{yx}=\mu\frac{\partial u}{\partial y}$) and an isotropic normal stress (i.e., $\sigma_{xx}=\sigma_{yy}=-p$), the set of Eqs.~(\ref{eq_moment}), (\ref{eq_pyy}), and (\ref{eq_cont2}) gives the Reynolds equation~\cite{Reynolds1886} to determine the pressure distribution, i.e., $p(x)$, along the thin fluid film under the nonslip boundary condition.
Even under the slip boundary condition, one can obtain a Reynolds equation-like model that determines the pressure distribution along a thin fluid film (see the Appendix \ref{app:slip}).
Below, we designate Eqs.~(\ref{eq:reynolds_slip}) and (\ref{eq:C_reynolds_slip}) modified Reynolds equations.

However, for complex fluids (e.g., colloids, polymers, liquid crystals, glasses, etc.), the constitutive relationships and boundary conditions are not generally known.
Thus, one cannot solve Eq.~(\ref{eq_moment}) analytically, and the Reynolds equation-like model cannot be obtained.

\subsection{Multiscale modeling}
In our strategy, instead of using any constitutive relations or macroscopic boundary conditions, we solve Eq.~(\ref{eq_moment}) using local MD cells.
The local MD cells are sparsely located at positions $x=x_i$ with a uniform interval $\varDelta x=L/I$, i.e., $x_i=i \varDelta x$ ($i=0,\cdots,I$), where $I$ is the number of intervals (see Fig.~\ref{fig:geom}).
The local MD cells are synchronized at a particular time interval to satisfy the macroscopic continuity equation (\ref{eq_cont2}).
Thus, one can obtain the cross-sectional velocity profiles in each MD cell and the pressure distribution along the channel, which simultaneously satisfy Eqs.~(\ref{eq_moment}), (\ref{eq_pyy}), and (\ref{eq_cont2}).

\red{
The interval $\varDelta x$ determines the streamwise resolution of the channel geometry and the local flow and stress fields, which may generally vary along the streamwise direction.
Accordingly, $\varDelta x$ should be chosen sufficiently small to resolve these variations while retaining the computational advantage of sparsely distributed MD cells.
}

We rewrite Eq.~(\ref{eq_moment}) using the first normal stress difference $\delta\sigma=\sigma_{xx}-\sigma_{yy}$ as follows to introduce the SMD scheme:
\begin{equation}\label{eq:momentum}
    \rho\frac{\partial u}{\partial t}
    =\frac{\partial \sigma_{yy}}{\partial x}
    +\frac{\partial \sigma_{yx}}{\partial y}
    +\frac{\partial \delta\sigma}{\partial x}.
\end{equation}
We note that at each local position $x=x_i$, $\frac{\partial \sigma_{yy}}{\partial x}$ is constant in $y$ (because of Eq.~(\ref{eq_pyy})), but the normal stress difference $\delta\sigma$ may vary in $y$.
We discretize Eq.~(\ref{eq:momentum}) at $x=x_i$ ($i=1,\cdots,I-1$) as follows:
\begin{equation}\label{eq_ui}
    \rho_i\frac{\partial u_i}{\partial t}
    =F_i
    +\frac{\partial \sigma_{yx,i}}{\partial y}
    +\frac{{\delta\sigma}_{i+1}-\delta\sigma_{i-1}}{2\varDelta x},
\end{equation}
where $F_i=\left.\frac{\partial \sigma_{yy}}{\partial x}\right|_{x=x_i}$ \red{represents the macroscopic streamwise gradient of the normal stress at $x=x_i$} and is uniform in $y$. 
\red{In the SMD implementation described below, this stress-gradient contribution is imposed on the local MD cell as a body-force-like numerical driving term.
It should not be interpreted as an additional physical external force generated, for example, by gravity or electromagnetic fields.}

At $x=x_0$ and $x=x_I$, the last term on the right-hand side of Eq.~(\ref{eq_ui}) is replaced with $(\delta\sigma_{1}-\delta\sigma_{0})/\varDelta x$ and $(\delta\sigma_{I}-\delta\sigma_{I-1})/\varDelta x$, respectively.

For the boundary conditions for the inlet and outlet of the channel at $x=0$ and $x=L$, we consider
\begin{equation}\label{eq_bcx}
\int_0^L\frac{\partial \sigma_{xx}}{\partial x} dx=\varDelta P,
\end{equation}
where $\varDelta P$ is the difference in pressure between the inlet and outlet of the channel.
By approximating the integration of Eq.~(\ref{eq_bcx}) with the trapezoidal rule, we obtain the following constraint on $\{F_i\}$, i.e.,
\begin{equation}\label{eq:const_F}
    \varDelta x\left[\frac{F_0+F_I}{2}+\sum_{i=\red{1}}^{I-1}F_i\right]
    +\overline{\delta\sigma_I}-\overline{\delta\sigma_0}=\varDelta P,
\end{equation}
where $\overline{\delta\sigma_{0,I}}$ is the spatial average of $\delta\sigma_{0,I}$ with respect to $y$.

We solve Eq.~(\ref{eq_ui}) using local MD cells, where the uniform \red{numerical driving} forces $F_i$ and the spatially varied \red{numerical driving} forces corresponding to the last term of Eq.~(\ref{eq_ui}) are subjected.

\subsection{MD cell configuration}
Rectangular MD cells $C_i$ ($i=0,\cdots,I$) are sparsely located at $x=x_i$ with a constant interval $\varDelta x$ along the channel (see Fig. \ref{fig:geom}).
The volume of the fluid region in the MD cell $C_i$ is $V_i=l_\mathrm{MD}^2\times h_i$, where $l_\mathrm{MD}$ is the side length of the MD cell and $h_i=h(x_i)$ is the channel height in $C_i$.
In each MD cell, the periodic boundary conditions are applied in the $x$ and $z$ directions, while the upper and lower walls \red{are represented by atomistic wall particles occupying 
the regions} $y=[h_i,h_i+W]$ and $y=[-W,0]$, respectively.
\blue{For each MD cell, the initial structure was prepared by randomly distributing the fluid particles, performing an energy minimization, and subsequently equilibrating the system in the NVT ensemble using a Langevin thermostat for a time much longer than the characteristic relaxation time of the fluid. Consequently, the temperature and density are uniform across all MD cells.
In the SMD scheme, }the wall temperature is fixed at a constant $T^w$ using a Langevin thermostat, while the fluid temperature is not controlled by any artificial thermostats but by the wall temperature, as in typical experimental systems.
The average number density of fluid molecules in each MD cell is constant and denoted as $\bar \rho$.
Thus, the total number of fluid molecules in the MD cell $C_i$ is $N_i=\bar \rho V_i$.

We calculate the steady-state solution of Eq.~(\ref{eq_ui}) $u_i(y)$ using the local MD cell $C_i$.
The channel height $h_i$ is divided into $N_\mathrm{bin}$ bins with a uniform height $\varDelta h_i=h_i/N_\mathrm{bin}$ to calculate the spatial distributions of macroscopic quantities in each MD cel, and in each bin, the local macroscopic quantities are calculated as follows:
\begin{subequations}\label{eq:macro_md}
\begin{align}
    \rho_i^l&=\frac{1}{|\varDelta V^l_i|}\sum_{n=1}^{N_i}\int_{\varDelta V_i^l} \delta(r_{y,n}-r_y)d\boldsymbol{r},\label{eq:rho_l}\\
    \rho_i^l u_i^l&=\frac{1}{|\varDelta V^l_i|}\sum_{n=1}^{N_i}\int_{\varDelta V_i^l} \dot r_{x,n}\delta(r_{y,n}-r_y)d\boldsymbol{r},\\
\sigma_{\alpha\beta,i}^l
&=\frac{1}{|\varDelta V_i^l|}
\sum_{n=1}^{N_i}
\int_{\varDelta V_i^l}\delta(r_{y,n}-r_y)
\left[(\dot r_{\alpha,n}-u_i^l\delta_{x\alpha})(\dot r_{\beta,n}-u_i^l\delta_{x\beta})+r_{\alpha,n}f_{\beta,n}\right]
d\boldsymbol{r},
\end{align}
\end{subequations}
where $\delta(x)$ is the Dirac delta function, the subscript $n\in\{1,\cdots,N_i\}$ denotes the molecular ID number in the MD cell $C_i$,
the superscript $l\in\{1,\cdots,N_\mathrm{bin}\}$ denotes the bin index, $\varDelta V_i^l=[0,l_\mathrm{MD}]\times[(l-1)\varDelta h_i,l\varDelta h_i]\times[0,l_\mathrm{MD}]$ and $|\varDelta V_i^l|$ denote the region and volume, respectively, of the $l$th bin in the MD cell $C_i$, $r_{\alpha,n}$ denotes the position of the $n$th molecule, and $f_{\alpha,n}$ denotes the force acting on that molecule due to interactions with surrounding molecules.

\red{
Although the prescribed mean number density $\bar{\rho}$ is specified uniformly among the local MD cells in each simulation case, the local bin-wise density, i.e., $\rho_i^l$ calculated by Eq.~(\ref{eq:rho_l}), may vary in the cross-sectional direction because of molecular-scale fluid--wall interactions, such as density layering near the solid walls. 
This cross-sectional density inhomogeneity is retained in the MD cells to accurately evaluate local velocities and stresses.
It should not be interpreted as a macroscopic density variation in the streamwise direction.
Indeed, the time-averaged wall-normal velocity in each local MD cell vanishes within statistical fluctuations. 
Therefore, the density inhomogeneity in the cross-sectional direction represents a stationary molecular-scale inhomogeneity near the walls, rather than a transported density field, and does not contradict the incompressibility condition imposed at the macroscopic level.
}

Since $F_i$ in Eq.~(\ref{eq_ui}) is identified as a uniform \red{numerical driving} force per unit volume in the fluid region, each fluid molecule in the $l$th bin $\varDelta V^l_i$ is driven by the \red{numerical driving} force
\begin{equation}\label{eq:calF}
    \mathcal{F}_i^l=\frac{F_i}{\rho_i^l},
\end{equation}
in the $x$ direction.
In addition to $\mathcal{F}_i^l$, the \red{numerical driving} force corresponding to the last term of Eq.~(\ref{eq_ui}), which is denoted as $\mathcal{G}_i^l$ in the following equation, is also needed.
The calculation of $\mathcal{G}_i^l$ is explained later.
We implement the MD simulations under the \red{numerical driving} force fields in each MD cell and calculate the local fluxes $\{\mathcal{M}_0, \cdots, \mathcal{M}_I\}$ at the cross-section at $x=x_0, \cdots, x_I$ according to Eq.~(\ref{eq_flux}), i.e.,
\begin{equation}\label{eq:integM}
    \mathcal{M}_i=\varDelta h_i\sum_{l=1}^{N_\mathrm{bin}} u_i^l,\quad (i=0,\cdots,I).
\end{equation}

\subsection{Synchronization of local MD cells}
\begin{figure}[t]
    \centering
    \includegraphics[width=0.6\linewidth]{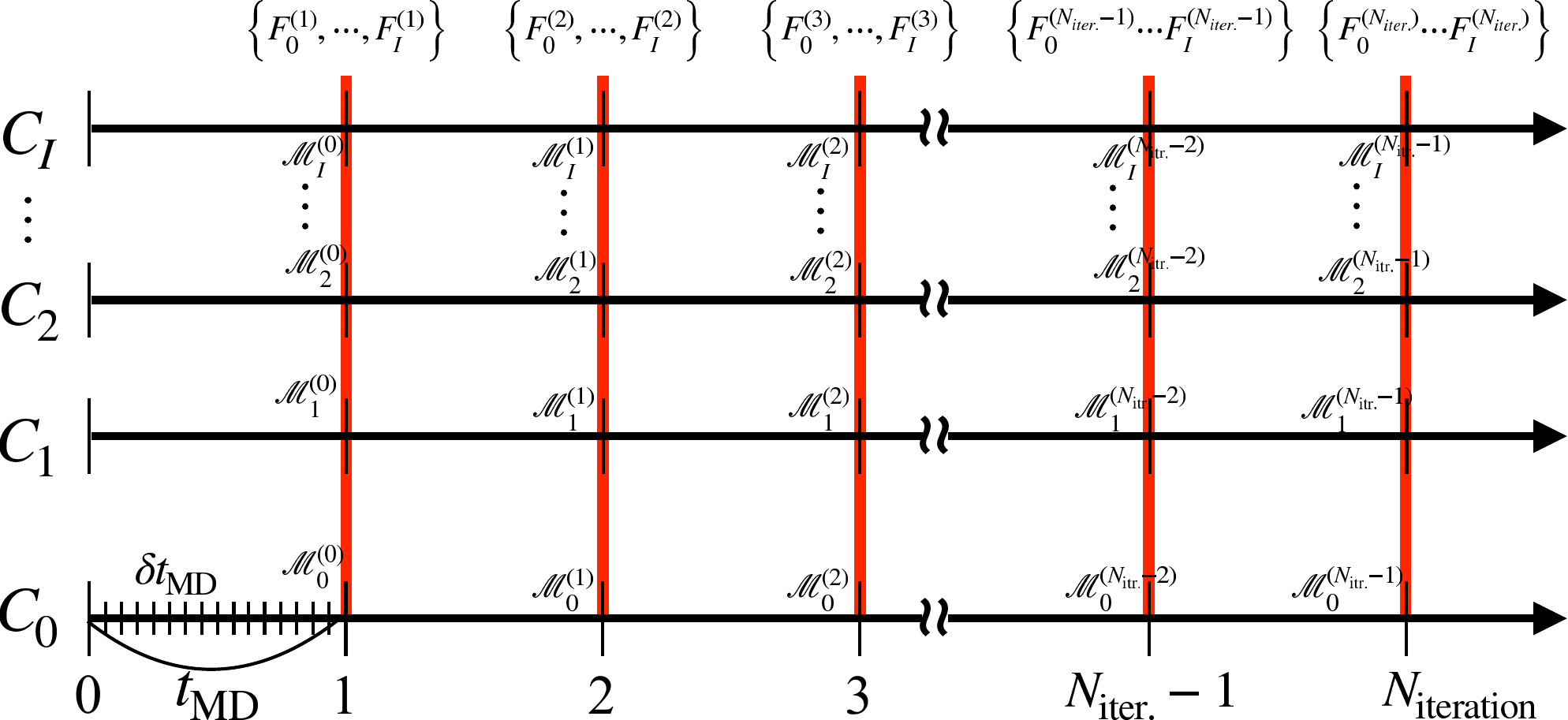}
    \caption{
    The synchronization scheme for local MD cells $\{C_0,\cdots,C_I\}$ proceeds as follows. 
    Given the \red{numerical driving} forces $\{F_0^{(k)},\cdots,F_I^{(k)}\}$ at iteration step $k$, independent MD simulations are performed in each MD cell over a time interval $t_\mathrm{MD}$ with the time-step size $\varDelta t_\mathrm{MD}$.
    After the MD runs are completed, the local fluxes obtained in each MD cell $\{\mathcal{M}_0^{(k)},\cdots,\mathcal{M}_I^{(k)}\}$ are exchanged among local MD cells.
    Finally, the \red{numerical driving} forces $\{F_0^{(k+1)},\cdots,F_I^{(k+1)}\}$ are updated so that the macroscopic continuity equation along the channel is satisfied.
    }
    \label{fig:scheme_smd}
\end{figure}

Our basic strategy is to find the set of \red{numerical driving} forces $\{F_0,\cdots,F_I\}$ such that the local fluxes $\{\mathcal{M}_0,\cdots,\mathcal{M}_I\}$ satisfy the continuity equation (\ref{eq_cont2}).
We achieved this goal using a simple iteration scheme, which is illustrated in
Fig.~\ref{fig:scheme_smd}.
At each iteration step, MD simulations are performed in every MD cell over a time interval $t_\mathrm{MD}$ under the \red{numerical driving} forces $\mathcal{F}^{l,(k)}_i$ and $\mathcal{G}_i^{l,(k)}$.
Hereafter, the superscript ``$(k)$'' denotes the $k$th iteration step.
The distributions of cross-sectional velocities $u_i^{l,(k)}$ and normal stress differences $\delta \sigma_i^{l,(k)}$ ($l=1,\cdots,N_\mathrm{bin}$) in the MD cell $C_i$ are calculated by time-averaging the instantaneous quantities obtained using Eq.~(\ref{eq:macro_md}) for the time period $t_\mathrm{MD}$.

First, we set the initial values of the \red{numerical driving} forces $\{{F}_0^{(0)},\cdots,{F}_I^{(0)}\}$ and implement the MD simulations with the given \red{numerical driving} forces in each MD cell for the time interval $t_\mathrm{MD}$.
Initially, we set $\mathcal{G}_i^{(0)}(y)=0$.
Then, the local fluxes $\mathcal{M}_i^{(0)}$ ($i=0,\cdots,I$) are calculated by integrating the cross-sectional velocities $\{u_i^{1,(0)},\cdots,u_i^{N_\mathrm{bin},(0)}\}$ in each MD cell using Eq.~(\ref{eq:integM}).

Once we obtain the local MD fluxes $\{\mathcal{M}_0^{(k)},\cdots,\mathcal{M}_I^{(k)}\}$ at iteration step $k$, we calculate the reference local fluxes $\{M_0^{(k)},\cdots,M_I^{(k)}\}$ that satisfy the continuity equation (\ref{eq_cont2}) using the following scheme:
\begin{subequations}\label{eq_Mi}
\begin{align}
&M_0^{(k)}=\overline{\cal M}^{(k)}+\frac{1}{L}\int_0^L\int_0^x[V_\mathrm{u}(\xi)-U_\mathrm{u}(\xi)h'_\mathrm{u}(\xi)]d\xi\,dx,\\
&M_i^{(k)}=M_{i-1}^{(k)}+\varDelta x\left[V_\mathrm{u}
-U_\mathrm{u}(x_i)h'_\mathrm{u}(x_i)
\right],\quad (i=1,\cdots,I),
\end{align}
\end{subequations}
where $\overline{\cal M}^{(k)}$ is the spatial average of the local MD fluxes along the channel, which is calculated using the trapezoidal rule as
\begin{equation}\label{eq:overline}
\overline{\cal M}^{(k)}=\frac1I\left[\frac{{\cal M}_0^{(k)}+{\cal M}_I^{(k)}}{2}+\sum_{i=1}^{I-1}{\cal M}_i^{(k)}\right].
\end{equation}
Afterward, we update the uniform \red{numerical driving} forces as follows:
\begin{subequations}\label{eq:updateF}
\begin{align}
F_i^{(k')}&=F_i^{(k)}+c(M_i^{(k)}-{\cal M}_i^{(k)}),\quad (i=0,\cdots,I),\label{eq:updateFa}\\
F_i^{(k+1)}&=F_i^{(k')}-\overline{F}^{(k')}+\frac{\varDelta P\red{+}\overline{\delta\sigma_0^{(k)}}-\overline{\delta\sigma_I^{(k)}}}{L},\quad (i=0,\cdots,I),\label{eq:updateFb}
\end{align}
\end{subequations}
where $c$ is an arbitrary positive constant and $\overline{F}^{(k')}$ is the spatial average of local MD \red{numerical driving} forces along the channel, which are calculated using the trapezoidal rule in Eq.~(\ref{eq:overline}).
Equation (\ref{eq:updateFb}) is required to satisfy Eq.~(\ref{eq:const_F}).

As mentioned below in Eq.~(\ref{eq:calF}), in addition to $F_i^{(k+1)}$, we need the cross-sectional distributions of the \red{numerical driving} forces ${\cal G}^{(k+1)}_i(y)$ in each MD cell, which are calculated as follows:
\begin{equation}
\begin{split}\label{eq_Gi}
&{\cal G}_i^{l,(k+1)}=\frac{\delta\sigma_{i+1}^{l,(k)}-\delta\sigma_{i-1}^{l,(k)}}{2\rho_i^l \varDelta x},\quad (i=1,\cdots,I-1),\\
&{\cal G}_{0}^{l,(k+1)}=\frac{\delta\sigma_{1}^{l,(k)}-\delta\sigma_{0}^{l,(k)}}{\rho_i^l \varDelta x},\quad
{\cal G}_{I}^{l,(k+1)}=\frac{\delta\sigma_{I}^{l,(k)}-\delta\sigma_{I-1}^{l,(k)}}{\rho_i^l \varDelta x},
\end{split}
\end{equation}
where ${\delta\sigma}_i^{l,(k)}=\sigma_{xx,i}^{l\,(k)}-\sigma_{yy,i}^{l\,(k)}$ is the normal stress difference calibrated in the $l$th bin of the MD cell $C_i$.
We repeat the overall iteration until the \red{numerical driving} forces $F_i^{(k)}$ ($i=0,\cdots,I$) converge as the local MD fluxes $\mathcal{M}_i^{(k)}$ ($i=0,\cdots,I$) satisfy the continuity equation (\ref{eq_cont2}).

\subsection{Implementation of the SMD method}
As illustrated in Fig.~\ref{fig:scheme_smd}, at each iteration step, MD simulations are performed in every MD cell over the time interval $t_\mathrm{MD}$, where MPI parallel computations are performed using the LAMMPS software package\cite{LAMMPS}.
To update these \red{numerical driving} forces, however, the cross-section fluxes $\mathcal{M}_i^{(k)}$ and the normal stress differences $\delta\sigma_i^{l,(k)}$ must be exchanged among different MD cells at every iteration step.
We construct a nested parallelization scheme employing the LAMMPS-C interface library to enable this hierarchical parallel computation\cite{2025_IOP_OY}.

\begin{figure}
    \centering
    \includegraphics[width=0.25\linewidth]{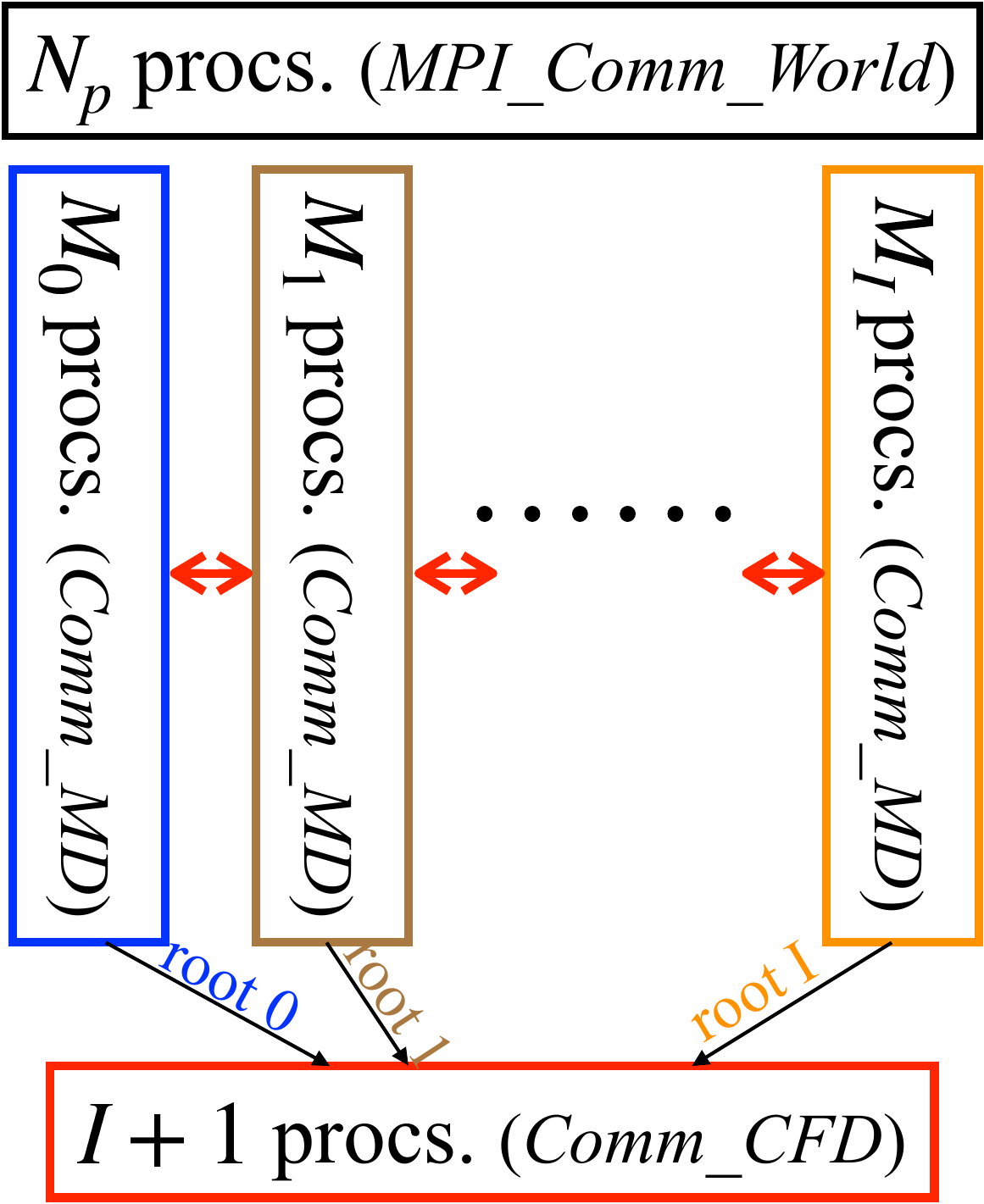}
    \caption{
    Schematic of the nested parallelization scheme of the SMD method.
    MD simulations in each MD cell are carried out independently within their corresponding sub-communicators (\textit{Comm\_MD}s).
    The exchange of local fluxes among MD cells is performed through another sub-communicator, \textit{Comm\_CFD}, which consists of the root rank of each \textit{Comm\_MD}.
    }
    \label{fig:comm_smd}
\end{figure}
Figure \ref{fig:comm_smd} illustrates the nested parallelization scheme.
We first initialize the global MPI communicator, \textit{MPI\_Comm\_World}, with $N_p$ processes, where $N_p$ denotes the total number of processes.
Each process is then assigned a local MD cell $C_i$ ($i=0,\cdots,I$), and \textit{MPI\_Comm\_World} is subsequently partitioned into $I+1$ subcommunicators, which are referred to as \textit{Comm\_MD}.
The number of processes assigned to MD cell $C_i$ is denoted by $M_i$, such that $N_p=\sum_{i=0}^IM_i$.
In addition, we construct another subcommunicator, \textit{Comm\_CFD}, which consists of the root rank of each \textit{Comm\_MD}.

\section{Validation testing}\label{sec:validation}
We consider a Lennard--Jones (LJ) fluid and compare the SMD results with those obtained from the Reynolds equation, both with and without slip boundary conditions, to validate the present SMD method.
The LJ fluid exhibits Newtonian behavior (i.e., constant viscosity) except in a very large high-shear-rate regime.
\red{
Although the present SMD method is formulated for incompressible flows, the low-density LJ-fluid cases are included only as validation tests. 
The purpose is to examine whether the SMD scheme can reproduce the theoretical solution for the slip flows of incompressible Newtonian fluid under prescribed mean densities, rather than to claim that the present formulation describes general compressible gas dynamics.
}

In each MD cell, both fluid and wall molecules are composed of LJ particles, which interact via the following potential:
\begin{equation}\label{eq:lj}
U_\mathrm{LJ}(r)=\left\{
\begin{array}{cc}
     4\varepsilon\left[\left(\frac{d}{r}\right)^{12}
     -\left(\frac{d}{r}\right)^{6}\right], &(r< r_c),  \\
     0, & (r\ge r_c),
\end{array}
\right.
\end{equation}
where $r$ is the distance between interacting particles, $d$ represents the diameter of the LJ particle, $\varepsilon$ is the energy unit, and $r_c$ is the cutoff parameter.
We fix the diameter of both fluid and wall particles equal to $d^f=d^w=d_0$, where the superscripts ``$f$'' and ``$w$'' represent the fluid and wall particles, respectively.
The energy parameter $\varepsilon$ and the cutoff parameter $r_c$ for the fluid--fluid and wall--wall interactions are also fixed as equal values, i.e., $\varepsilon^{ff}=\varepsilon^{fw}=\varepsilon_0$ and $r_c^{ff}=r_c^{ww}=2^\frac16 d_0$.
On the other hand, for the fluid--wall interaction, we consider three different types of models, as shown in Table \ref{tab:ljpara}.
Hereafter, we measure the space and time in units of $d_0$ and
$md_0^2/\varepsilon_0$, respectively.
Here, $m$ is the mass of the LJ particle, which is common in both the wall and the fluid.
The temperature $T$ is measured in units of $\varepsilon_0/k_B$.
For M1, all the particles interact with the truncated LJ potential, where no attraction force is considered in any of the fluid--fluid, wall--wall, or fluid--wall interactions.
For M2 and M3, the attraction force is considered in the fluid--wall interaction, which may affect slip flows at the wall boundary.
\begin{table}[t]
    \centering
    \begin{tabular}{c cc}
 \hline\hline
         &  $\varepsilon^{fw}$ & $r_c^{fw}$\\
         \hline
     M1   &  1 & $2^\frac16$\\
     M2  &  0.2 & 2.5\\
     M3 &  0.7 & 2.5\\
    \hline\hline
    \end{tabular}
    \caption{Three different types of models for the fluid--wall interaction: energy and cutoff parameters of the LJ potential (\ref{eq:lj}) used for the fluid--wall interaction. 
    }
    \label{tab:ljpara}
\end{table}

The wall particles are connected to the face-centered cubic
(fcc) lattice structure by a spring potential, and the temperature of the wall particles is kept constant at $T^w$ by the Langevin
thermostat algorithm \cite{1997_PRE_TKE,2019_PRE_NOY,Allen_Tildesley_2017,Evans_Morriss_2008}.
On the other hand, the fluid temperature is not artificially controlled by any thermostat algorithm.

The geometry of the channel is depicted in Fig.~\ref{fig:geom}.
We consider a wedge-shaped channel.
The channel height is set as follows:
$h_0=317.48$, $h_I=158.74$, $L=2000$, and $h'(x)=(h_I-h_0)/L$.
The base size of the rectangular MD cells is fixed at $l_\mathrm{MD}=31.75$, and the number of intervals between MD cells is set to $I=7$, i.e., $\varDelta x=285.7$.
Thus, the efficiency of the SMD simulation compared with that of the full MD simulation is represented as $\varDelta x/l_\mathrm{MD}=9$.
The thickness of the slab wall in each MD cell is set to $W=3.97$.

At each iteration step, MD simulations are performed over a time interval $t_\mathrm{MD}=5$ using a time step size of $\varDelta t_\mathrm{MD}$=0.005.
Thus, each MD run consists of $N_\mathrm{mdrun}=1000$ time steps.
We set the iteration number to $N_\mathrm{iteration}=10000$ and evaluate the steady-state force distribution along the channel, as well as the cross-sectional velocity profiles at the cross sections in each MD cell, by taking time averages over the final half of the simulation, i.e., over the interval of $t_\mathrm{MD}N_\mathrm{iteration}/2$ \red{(i.e., $5\times 10^6$ MD steps)}.
We set the coefficient $c$ in Eq.~(\ref{eq:updateFa}) to $c=\bar\rho/h_i^3$.
The convergence of the iteration scheme is discussed in Sec.~\ref{sec:convergence}.

The wall temperature $T^w=1$ and density $\rho^w=1$ are fixed.
The temperature of the fluid is initially set at $T=1$ and remains nearly constant in the subsequent simulations, as the wall temperature is kept constant and the viscous heating is very small in the subsequent simulations.

\subsection{Pressure-driven flows}
\begin{figure}[t]
    \centering
    \includegraphics[width=0.7\linewidth]{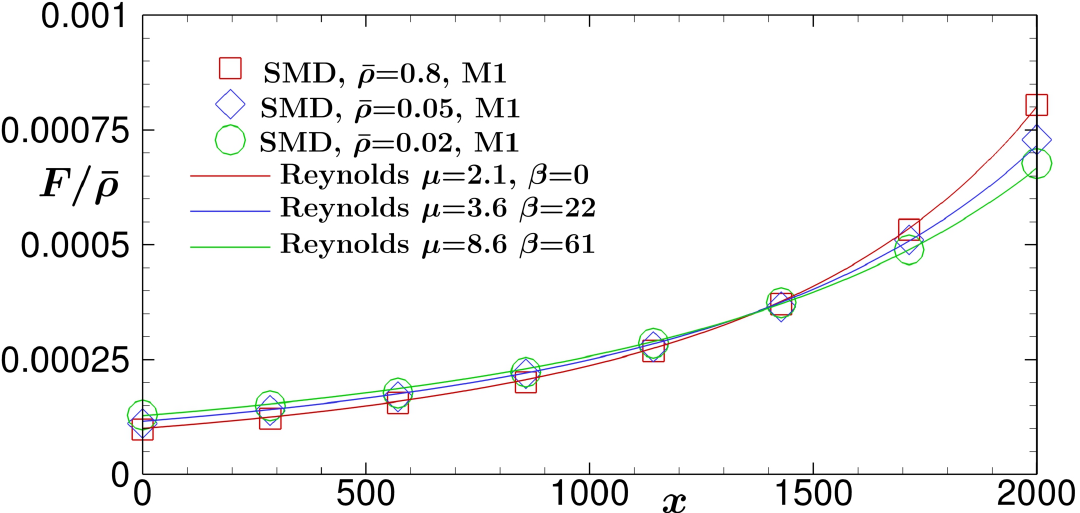}
    \caption{Force distributions along the channel for the pressure-driven flows with $\varDelta P/\bar \rho=0.6$. The red triangles show the SMD results for the liquid flow (i.e., $\bar \rho=0.8$), the blue downward triangles show those for the \red{low-density LJ fluid} flow (i.e., $\bar \rho=0.05$), and the green squares show those for the \red{very-low-density LJ fluid} flow (i.e., $\bar \rho=0.02$). In all cases, the fluid--wall interaction model is set to M1 in Table~\ref{tab:ljpara}. The solid lines show the results obtained using the modified Reynolds equation, i.e., Eqs.~(\ref{eq:reynolds_slip}) and (\ref{eq:C_reynolds_slip}), where the viscosity $\mu$ and slip length $\beta$ are calibrated from the SMD results using Eq.~(\ref{eq:calib_viscos}).}
    \label{fig:reynolds_delp}
\end{figure}
In this subsection, we examine pressure-driven flows in three distinct \red{prescribed mean densities of the LJ fluid}: a liquid at $\bar \rho=0.8$, a \red{low-density LJ fluid} at $\bar \rho=0.05$, and a \red{very-low-density LJ fluid} at $\bar \rho=0.02$.
The difference in pressure between the inlet and outlet of the channel is set to $\varDelta P/\bar \rho=0.6$.
In all cases, the fluid--wall interaction is fixed to Model M1, as listed in Table~\ref{tab:ljpara}.

Figure \ref{fig:reynolds_delp} shows the force distributions along the channel.
Here, the results of the SMD simulations are compared with those of the modified Reynolds equation, i.e., Eqs.~(\ref{eq:reynolds_slip}) and (\ref{eq:C_reynolds_slip}), where the viscosity $\mu$ and slip length $\beta$ are calibrated from the SMD results.
The SMD results agree well with those obtained using the modified Reynolds equation, i.e., Eqs.~(\ref{eq:reynolds_slip}) and (\ref{eq:C_reynolds_slip}), for all fluid densities.

\begin{figure}[t]
    \centering
    \includegraphics[width=0.8\linewidth]{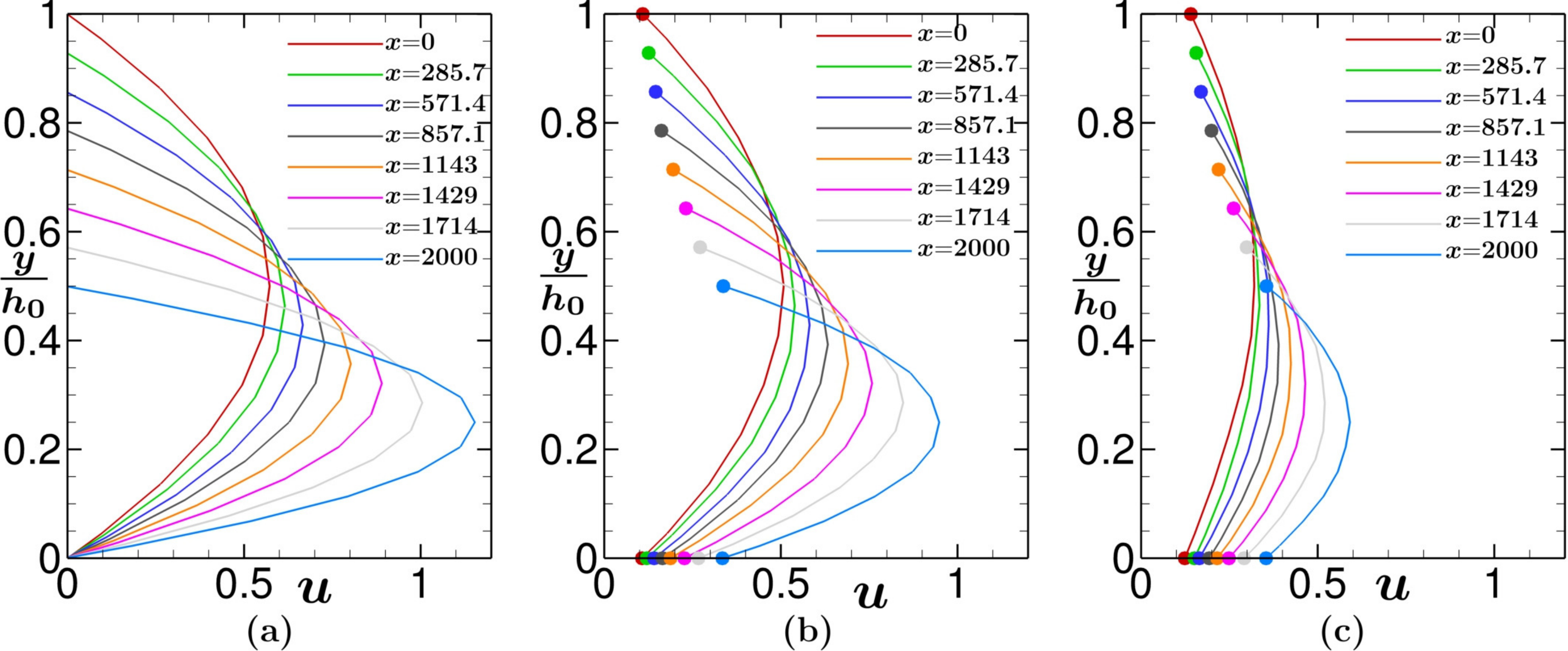}
    \caption{Velocity profiles of each MD cell for the pressure-driven flows with distinct fluid densities: $\bar \rho$=0.8 in (a), $\bar \rho$=0.05 in (b), and $\bar \rho$=0.02 in (c).
    The pressure difference between the inlet and outlet of the channel is set to $\varDelta P/\bar \rho=0.6$. 
    The circles in panels (b) and (c) denote the slip velocities at the wall boundaries.}
    \label{fig:vx_delp}
\end{figure}

\red{Figures \ref{fig:vx_delp}(a)--\ref{fig:vx_delp}(c)} show the velocity profiles in each MD cell.
Each velocity profile has a parabolic shape, as expected for a Newtonian fluid.
For the \red{low-density LJ fluid} flows (see Fig.~\ref{fig:vx_delp}(b) and (c)), significant slip velocities are observed at the wall boundaries.
The magnitude of the slip velocity increases as the channel height decreases.
This result occurs because the slip velocity is proportional to the velocity gradient at the boundary, i.e., $u_\mathrm{slip}=\beta\left.\frac{\partial u}{\partial y}\right|_\mathrm{wall}$, and the velocity gradient increases as the channel height decreases.

\begin{table}[t]
    \centering
    \begin{tabular}{c c c c }
    \hline\hline
         ($\varepsilon^{fw}$,$r_c^{fw}$) &$\bar \rho$&$\beta$ & $\xi$\\
         \hline
         M1 &0.8 &0 &--\\
         M1 &0.05 & 22& 4.50\\
         M1 &0.02 & 62& 11.25 \\
         M2 &0.05 & 10&4.50 \\
         M3 &0.05 & 2.4&4.50 \\
         \hline\hline
    \end{tabular}
    \caption{
    The slip length $\beta$ is calibrated from the velocity profiles in Figs.~\ref{fig:vx_delp} and \ref{fig:vx_vw}. The mean free path $\xi=(\sqrt{2}\pi\bar \rho{\sigma_0}^2)^{-1}$ for a hard-sphere gas is also presented for $\bar \rho=0.02$ and $\bar \rho=0.05$.
    The fluid-wall interaction parameters for M1--M3 are shown in Table~\ref{tab:ljpara}.   
    }
    \label{tab:slip}
\end{table}

The viscosity $\mu$ and slip length $\beta$ are calibrated as described below.
The local viscosity $\mu_i^l$ at the $l$th bin in the MD cell $C_i$ is computed from the local flow velocity $u_i^l$ and the shear stress $\sigma_{xy,i}^l$ obtained from Eq.~(\ref{eq:macro_md}), as follows:
$$
\mu_i^l = \frac{\sigma_{xy,i}^l}{\dot{\gamma}_i^l}, 
\qquad l = 2,\dots,N_\mathrm{bin}-1,
$$
where the local shear rate is evaluated by calculating the central difference
$$
\dot{\gamma}_i^l = \frac{u_i^{l+1}-u_i^{l-1}}{2\delta h_i}.
$$
The viscosity $\mu_i$ of the MD cell $C_i$ is then evaluated using a shear-rate-weighted average,
$$
\mu_i =
\left(\displaystyle\sum_{l=2}^{N_\mathrm{bin}-1} 
\left|\dot{\gamma}_i^l\right| \mu_i^l\right)\Big/
\left(\displaystyle\sum_{l=2}^{N_\mathrm{bin}-1} 
\left|\dot{\gamma}_i^l\right|\right).
$$
Finally, the overall viscosity is obtained by averaging over all MD cells,
\begin{equation}\label{eq:calib_viscos}
\mu = \frac{1}{I}\sum_{i=1}^{I} \mu_i .
\end{equation}

The slip length is determined by extrapolating the velocity profile in the vicinity of the boundary to the position where the velocity vanishes.
We evaluate the slip lengths $\beta_i^{\mathrm{u}}$ and $\beta_i^{\mathrm{b}}$ at the upper and lower boundaries, respectively, in each MD cell $C_i$.
The overall slip lengths at the upper and lower boundaries are then obtained by taking a weighted average with respect to the slip velocity at the boundary, i.e.,
\begin{equation}
    \beta^{\mathrm{u,b}}=\left(\sum_{i=1}^I |v_{x,i}^{\mathrm{u,b}}| \, \beta_i^{\mathrm{u,b}} \right)\bigg/ \left(\sum_{i=1}^I |v_{x,i}^{\mathrm{u,b}}|\right), 
    \quad 
    \beta=\frac{\beta^{\mathrm{u}}+\beta^{\mathrm{b}}}{2},
\end{equation}
where $v_{x,i}^{\mathrm{u,b}}$ denote the slip velocities at the upper and lower walls, respectively, in the MD cell $C_i$.
The calibrated slip lengths $\beta$, obtained from Figs.~\ref{fig:vx_delp} and \ref{fig:vx_vw}, are summarized in Table~\ref{tab:slip}.
The slip length is several times longer than the mean free path for weak fluid--wall interaction cases (M1 and M2), which is consistent with expectations from the kinetic theory of gases~\cite{sone2007}.

\subsection{Wall-driven flows}
\begin{figure}[t]
    \centering
    \includegraphics[width=0.7\linewidth]{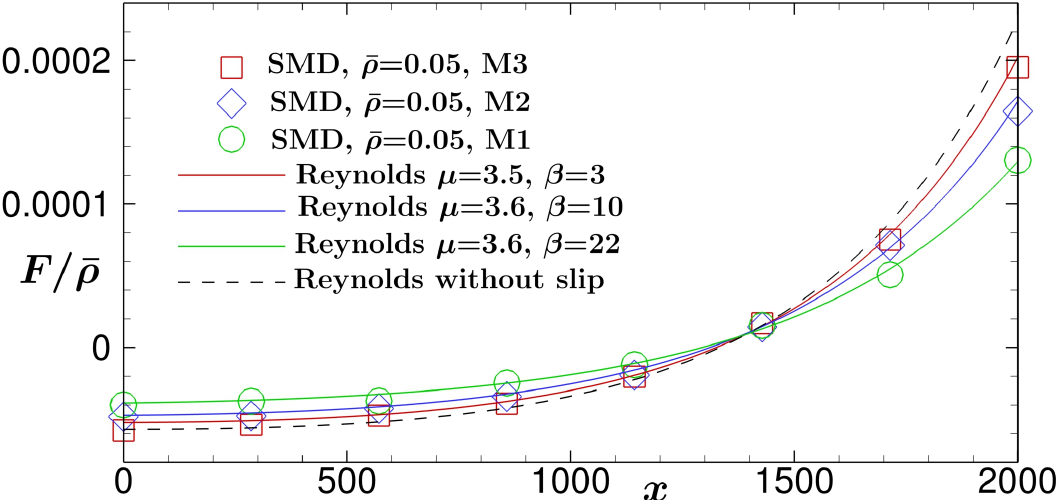}
    \caption{Force distributions along the channel for wall-driven flows with $U_b$=0.8 and $U_\mathrm{u}$=$V_\mathrm{u}$=0.
    The SMD results for three different fluid-wall interaction models (M1--\red{M3} in Table~\ref{tab:ljpara}) are shown: red squares for the moderate interaction case (M3), blue diamonds for the weak interaction case (M2), and green circles for the purely repulsive case (M1).
    See also the caption of Fig.~\ref{fig:reynolds_delp}.
    }\label{fig:reynolds_vw_rho005}
\end{figure}
\begin{figure}
    \centering
    \includegraphics[width=0.8\linewidth]{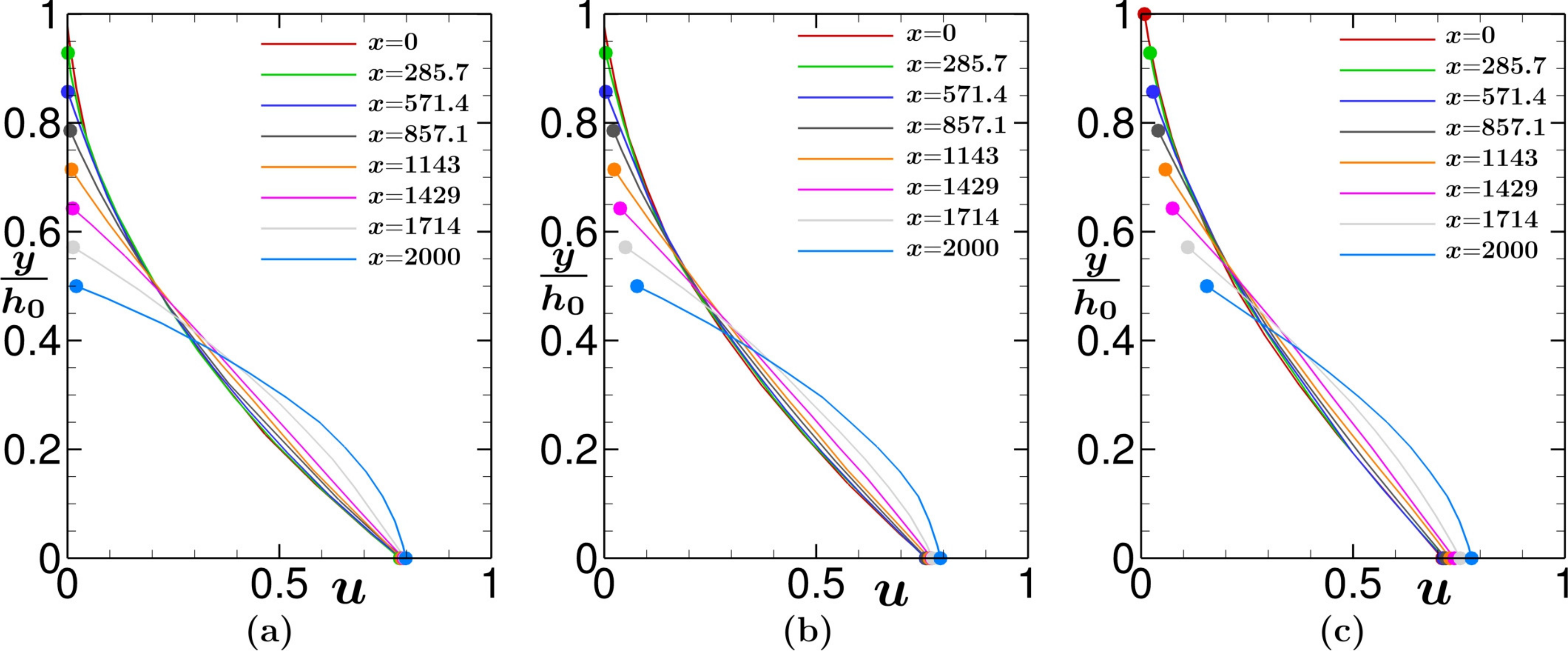}
    \caption{Velocity profiles of each MD cell for wall-driven flows for three different wall interaction models (M1--M3 in Table~\ref{tab:ljpara}): the moderate interaction case (M3) in (a), the weak interaction case (M2) in (b), and the purely repulsive case (M1) in (c).
    The wall velocities are set to $U_\mathrm{u}=V_\mathrm{u}=0$ and $U_\mathrm{b}=0.8$ and the pressure difference between the inlet and outlet of the channel is set to $\varDelta P/\rho=0$.
    The fluid density $\bar \rho=0.05$ is fixed.
    In each panels, the circles denote the slip velocities at the wall boundaries.}
    \label{fig:vx_vw}
\end{figure}

In this subsection, we examine wall-driven flows for different fluid--wall interactions, i.e., M1--M3 in Table \ref{tab:ljpara}, while the fluid density is fixed as $\bar \rho=0.05$.
The bottom wall slides with a velocity $U_\mathrm{b}=0.8$ while the upper wall is at rest, i.e., $U_\mathrm{u}$=$V_\mathrm{u}$=0.
The difference in pressure between the inlet and outlet of the channel is set to $\varDelta P$=0.

Figures~\ref{fig:reynolds_vw_rho005} and \red{\ref{fig:vx_vw}(a)--\ref{fig:vx_vw}(c)} present the force distributions along the channel and the velocity profiles in each MD cell, respectively.
For the relatively large fluid--wall interaction case (M3), the slip velocities at the walls remain very small (see Fig.~\ref{fig:vx_vw}(a)), whereas they become significantly larger for the purely repulsive case (M1) (see Fig.~\ref{fig:vx_vw}(c)).
The influence of the slip velocity on the force distribution along the channel is clearly observed in Fig.~\ref{fig:reynolds_vw_rho005}: when the slip velocity is high, the spatial variation in the force weakens. \blue{This behavior can be understood as follows. As the slip velocity increases, the wall-induced velocity gradient near the moving wall decreases, and consequently the shear stress exerted on the fluid by the wall also decreases. Accordingly, a smaller local force $F$, which is equivalent to minus the streamwise pressure gradient for a Newtonian fluid, is required to satisfy the local momentum balance. In addition, because $\Delta P=0$ in the present wall-driven problem, the streamwise integral of $F$ must vanish. Therefore, as the magnitude of $F$ decreases with increasing slip, the streamwise variation of $F$ becomes weaker.}

Overall, the SMD results show excellent agreement with those obtained from the modified Reynolds equation, Eqs.~(\ref{eq:reynolds_slip}) and (\ref{eq:C_reynolds_slip}), for all fluid--wall interaction models considered.
These findings indicate that the present SMD framework accurately captures the hydrodynamic behavior of simple LJ fluids, including slip flows at the boundaries.

\subsection{Convergence of the iteration scheme}\label{sec:convergence}
We examine the convergence of the iteration scheme (Eqs.~(\ref{eq_Mi})--(\ref{eq:updateF})) by varying the number of MD steps per iteration, i.e., $N_\mathrm{mdrun}$=$\{10,10^2,10^3,10^4,10^5\}$, while fixing the total number of MD steps as $N_\mathrm{mdrun}\times N_\mathrm{iteration}=2\times 10^7$.

\red{In a macroscopically steady state, the statistical uncertainty arising from molecular thermal fluctuations generally decreases with the total sampling time. 
Nevertheless, in the SMD method, even when the total number of MD steps is fixed, $N_\mathrm{mdrun}$ may affect how the statistical fluctuations enter the synchronization process.
When $N_\mathrm{mdrun}$ is small, the MD-cell-averaged quantities evaluated at each iteration are obtained over a short sampling period and are therefore strongly affected by thermal fluctuations.
Conversely, a large $N_\mathrm{mdrun}$ increases the interval between successive synchronization updates and reduces the total number of iterations.
Consequently, cell-to-cell inconsistencies may persist until the next synchronization update, leading to oscillatory adjustments among the local MD cells. 
To examine this trade-off, we compare the evolution of $Err^{(k)}$, defined below, for different values of $N_\mathrm{mdrun}$.}

\begin{figure}
    \centering
    \includegraphics[width=0.5\linewidth]{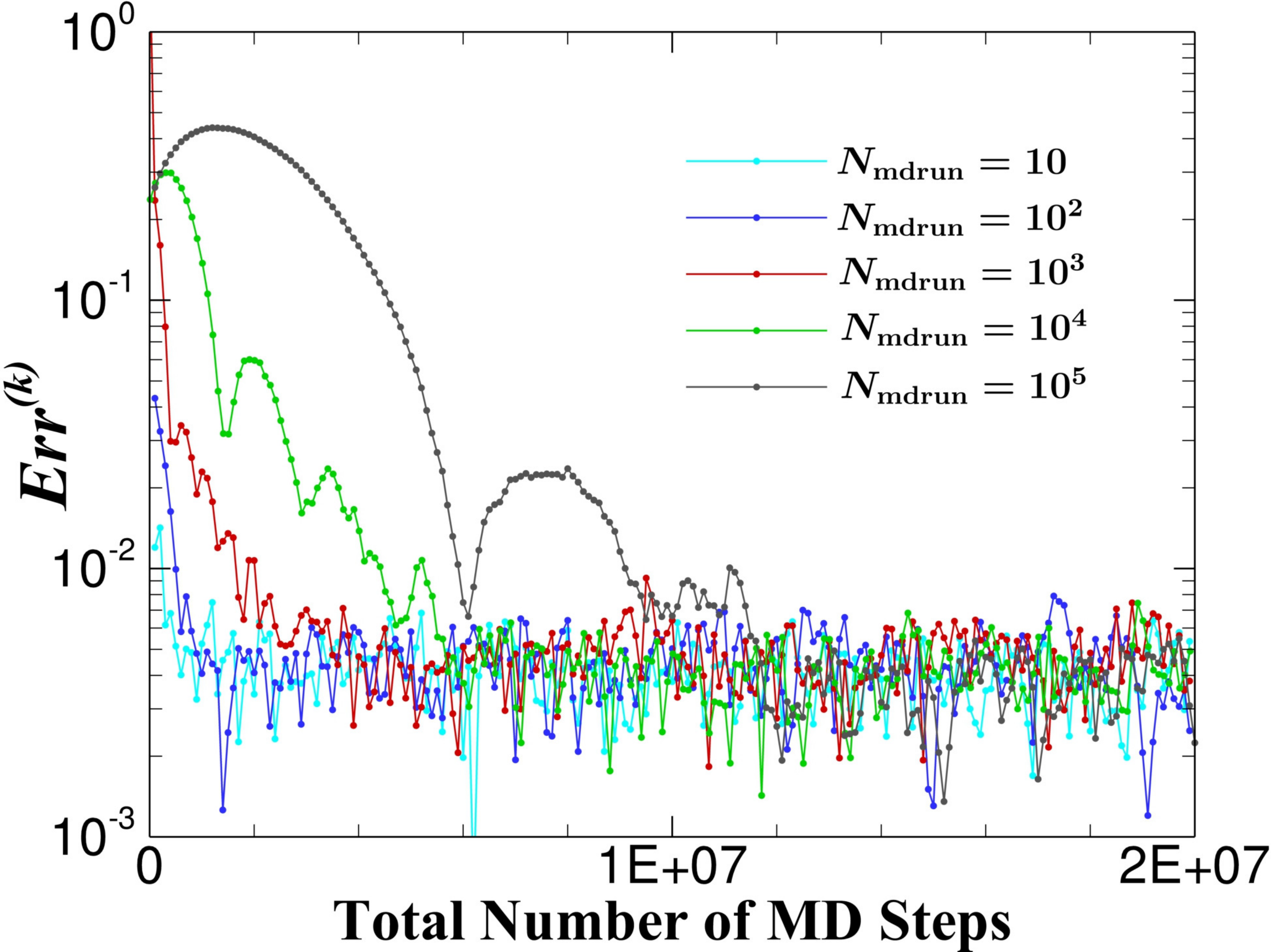}
    \caption{The convergence behaviors of Eq.~(\ref{eq:errk}) for different values of $N_\mathrm{mdrun}$ are plotted against the total number of MD steps $kN_\mathrm{mdrun}$.
    The pressure-driven flow with $\bar \rho$=0.8, $\varDelta P/\bar \rho$=0.6, and fluid-wall interaction M1 is computed.}
    \label{fig:conv}
\end{figure}

Figure \ref{fig:conv} shows the convergence behavior, which was quantified by the normalized error defined as
\begin{equation}\label{eq:errk}
    {Err}^{(k)}=\sqrt{\frac{1}{I+1}\sum_{i=0}^I\left(\frac{\mathcal{M}_i^{(k)}-M_i^{(k)
    }}{\overline{\cal M}^{(k)}}\right)^2},\quad (k=1,\cdots,N_\mathrm{iteration}).
\end{equation}
Here, $k$ denotes the iteration step, $\mathcal{M}_i^{(k)}$ is the local flux in the MD cell $C_i$, $M_i^{(k)}$ is the reference flux in the MD cell $C_i$ given by Eq.~(\ref{eq_Mi}), and $\overline{\cal M}^{(k)}$ is the spatial average of $\mathcal{M}_i^{(k)}$ along the channel given by Eq.~(\ref{eq:overline}).
\begin{figure}
    \centering
    \includegraphics[width=0.5\linewidth]{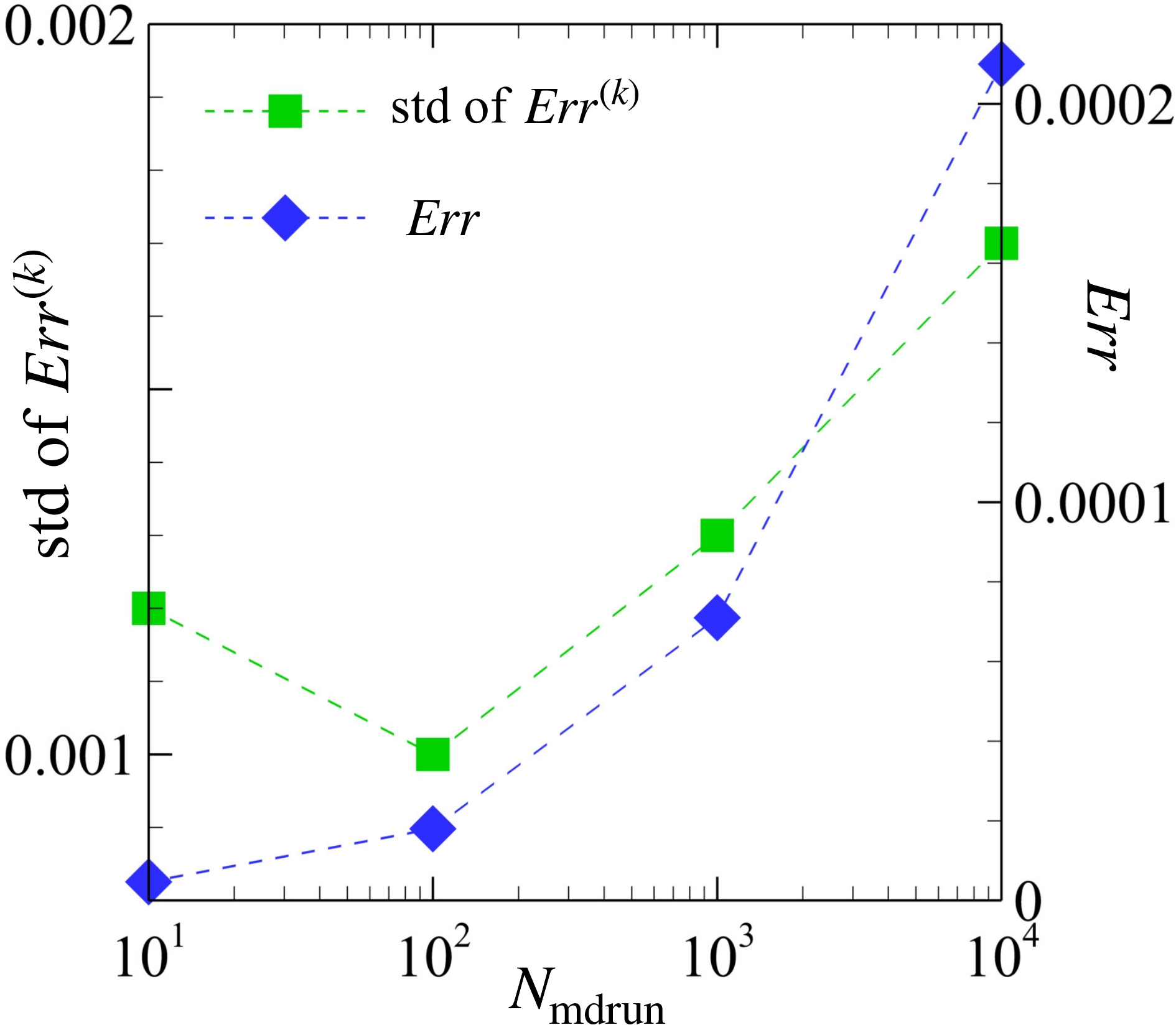}
    \caption{The standard deviation of $Err^{(k)}$ over the interval $k N_\mathrm{mdrun}\in[10^7,2\times 10^7]$ and the error estimate $Err$ of the time-averaged quantity $\mathcal{M}_i^{(k)}$ in $k N_\mathrm{mdrun}\in[10^7,2\times 10^7]$, which were obtained by replacing $\mathcal{M}_i^{(k)}$ with its time average, are shown for different values of $N_\mathrm{mdrun}$.}
    \label{fig:std_error}
\end{figure}
The convergence becomes slower as $N_\mathrm{mdrun}$ increases.
In all cases, $Err^{(k)}$ initially decreases to approximately $4\times 10^{-3}$, after which it appears to fluctuate around that level.
\red{
Remarkably, for large $N_\mathrm{mdrun}$, $Err^{(k)}$ exhibits the slow, oscillatory convergence discussed above.
}

The standard deviation of $Err^{(k)}$ over the interval $k N_\mathrm{mdrun}\in[10^7,2\times 10^7]$ and the error estimate \red{$Err$} of the time-averaged quantity $\mathcal{M}_i^{(k)}$ in the same interval, \red{obtained by replacing $\mathcal{M}_i^{(k)}$ with its time average,} are shown in Fig.~\ref{fig:std_error}.
\red{For large $N_\mathrm{mdrun}$, both the standard deviation of $Err^{(k)}$ and the error estimate $Err$ increase with $N_\mathrm{mdrun}$.
This behavior may be associated with the slow, oscillatory convergence caused by less frequent synchronization updates.
On the other hand, when $N_\mathrm{mdrun}$ is very small, the standard deviation increases because the quantities evaluated at each iteration are more strongly affected by thermal fluctuations.
These results clearly demonstrate the trade-off discussed above.
Consequently,}
an optimal value of $N_\mathrm{mdrun}$ exists around $N_\mathrm{mdrun}\sim 100$, which yields a convergent behavior with small fluctuations.

Notably, because MPI communications among different MD cells are required at every iteration step in the SMD implementation, the communication load increases as $N_\mathrm{mdrun}$ decreases, even when the total number of MD steps is fixed.
Considering both the total computational cost and the convergent behavior of the error, we chose $N_\mathrm{mdrun}$=1000 in Sec.~\ref{sec:validation}.

\section{Application to polymer lubrication flows}
We consider a Kremer--Grest (KG) model polymeric fluid composed of chains containing $N_b$ LJ beads.
The beads interact via the purely repulsive part of the LJ potential (\ref{eq:lj}), while consecutive beads within the same chain are connected by a finite extensible nonlinear elastic (FENE) potential, i.e.,
\begin{equation}\label{eq:fene}
    U_\mathrm{F}(r)=-0.5KR_0^2\ln\left(1-\frac{r}{R_0}\right),
\end{equation}
where the constants $K=30$ and $R_0=1.5$ are fixed.
In the subsequent simulations, the number of beads per chain is set to $N_b$=32, which corresponds to an unentangled polymer chain.

The channel walls are composed of LJ particles, as described in the previous section, with the wall temperature and density fixed at $T^w$=1 and $\rho^w=1$, respectively.
The initial temperature and average density of the model polymeric fluid are set to $T=1$ and $\bar \rho=0.8$, respectively.
The fluid--wall interaction corresponds to case M1 in Table~\ref{tab:ljpara}.

The channel geometry is set as $L=4000$, $h_0=317.48$, and $h_I=158.74$ in Fig.~\ref{fig:geom}.
The base size of the rectangular MD cells is fixed at $l_\mathrm{MD}=\red{31.75}$, and the number of intervals between MD cells is set to $I=7$.
Thus, the efficiency of the SMD simulation compared with that of the full MD simulation is represented as $\varDelta x/l_\mathrm{MD}$=18.

\begin{figure}
    \centering
    \includegraphics[width=0.7\linewidth]{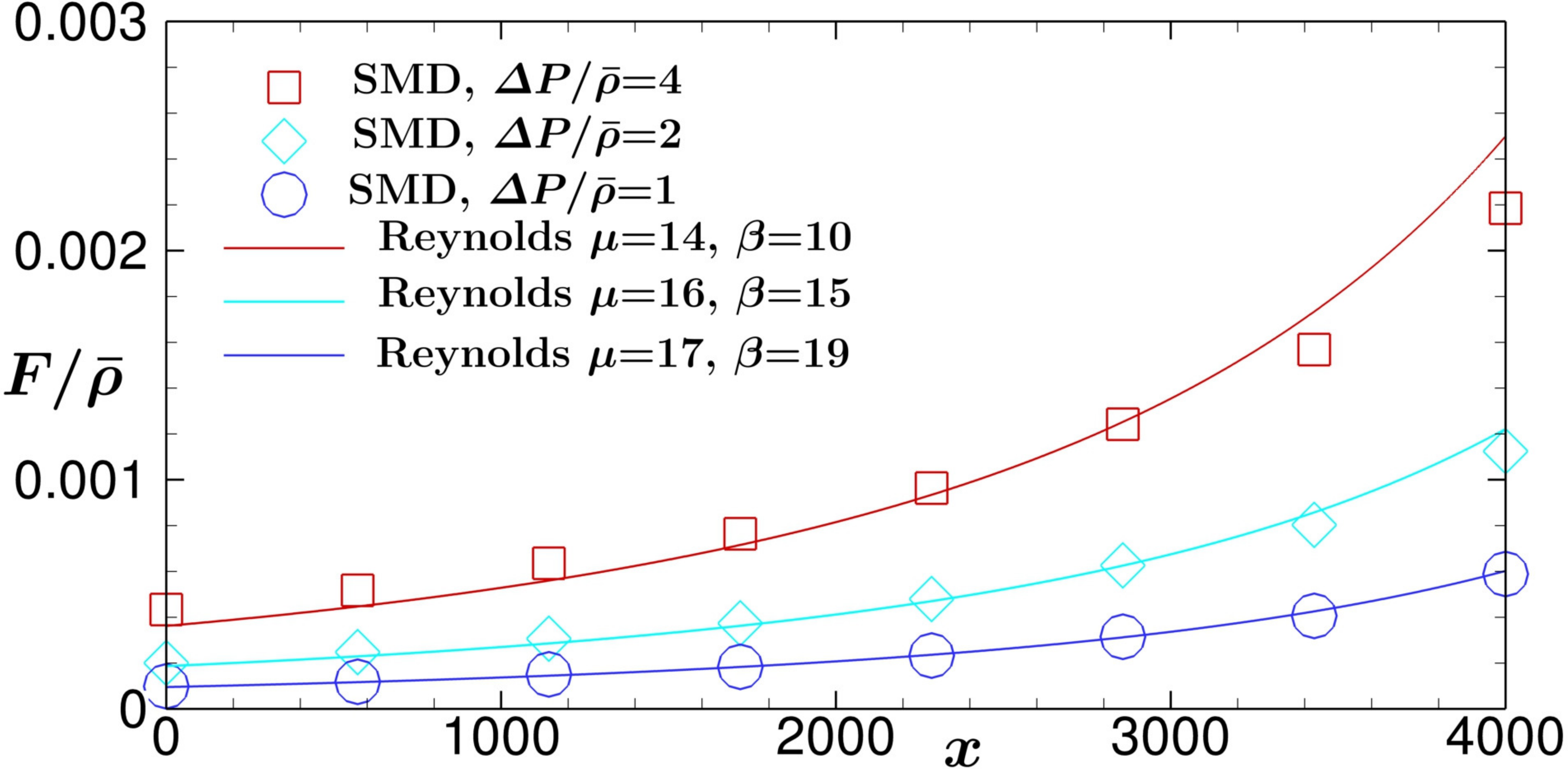}
    \caption{Force distributions along the channel for pressure-driven flows of a polymeric fluid with $\bar\rho$=0.8 and $N_b$=32.
    The symbols show the SMD results for different values of pressure difference $\varDelta P$ as indicated in the figure.
    The solid lines show the results obtained by the modified Reynolds equation, i.e., Eqs.~(\ref{eq:reynolds_slip}) and (\ref{eq:C_reynolds_slip}), where the viscosity $\mu$ and slip length $\beta$ are calibrated from the SMD results as Eq.~(\ref{eq:calib_viscos}). 
    }\label{fig:reynolds_nb32}
\end{figure}

Figure~\ref{fig:reynolds_nb32} shows the comparisons of the force distributions obtained using SMD and those obtained using the modified Reynolds equation (\ref{eq:reynolds_slip}) and (\ref{eq:C_reynolds_slip}).
In the Reynolds equation, the viscosity $\mu$ and slip coefficient $\beta$ are evaluated as described in the previous section (see Eq.~(\ref{eq:calib_viscos})).
A distinct deviation between the SMD results and those obtained using the modified Reynolds equation is observed at a large pressure difference of $\varDelta P/\bar\rho$=4.

This deviation arises from the non-Newtonian behavior of the polymeric fluid, specifically shear thinning.
The viscosity of the model polymeric fluid decreases once the shear rate exceeds a certain critical value.
Consequently, the modified Reynolds equation, which is based on the Newtonian fluid assumption (i.e., a constant viscosity), is no longer applicable to high-speed flows at large pressure differences.

\begin{figure}
    \centering
    \includegraphics[width=0.8\linewidth]{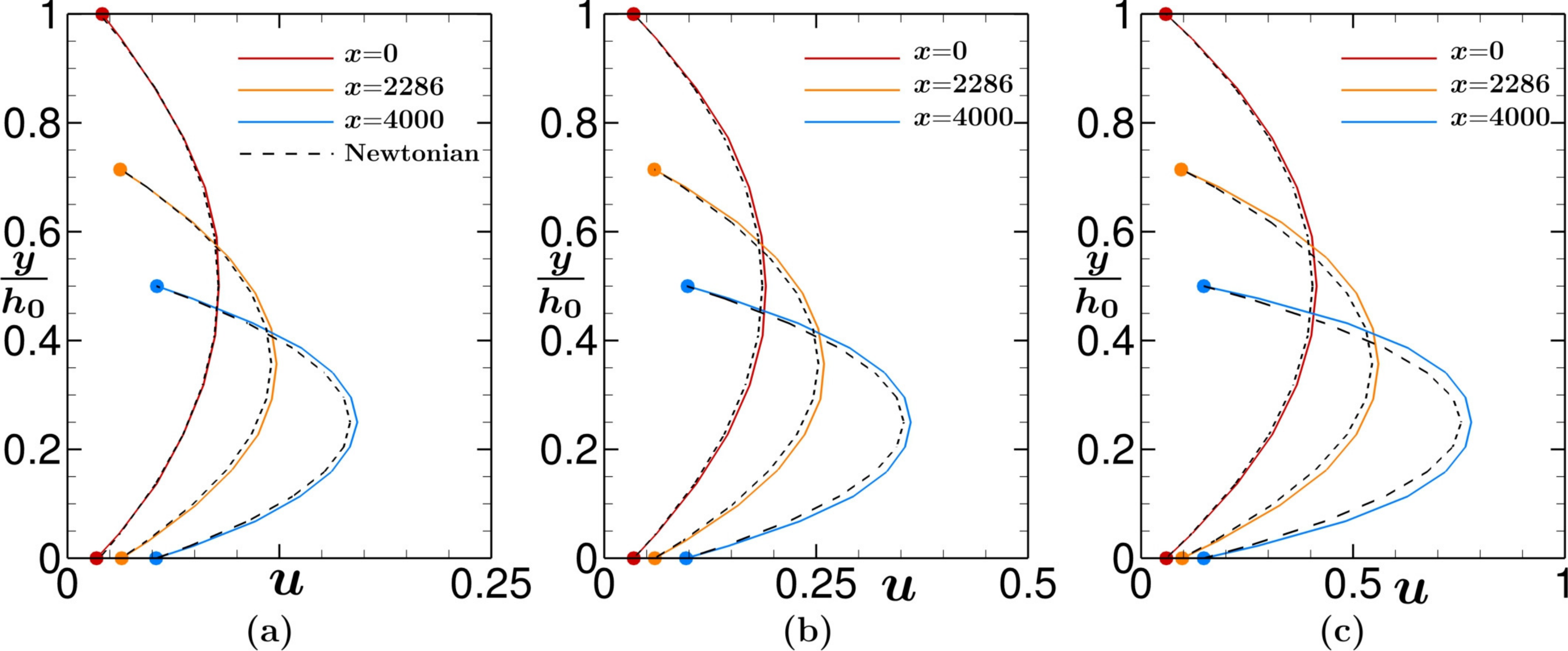}
    \caption{
   Comparison of velocity profiles between the model polymeric fluid (solid lines) and the corresponding Newtonian fluids (dashed lines).
    The left panel [(a)] shows the results for $\varDelta P/\bar\rho = 1$, the middle panel [(b)] for $\varDelta P/\bar\rho = 2$, and the right panel [(c)] for $\varDelta P/\bar\rho = 4$.
    The viscosity and slip velocity of the Newtonian fluids, evaluated from the SMD results, are indicated in the legend of Fig.~\ref{fig:reynolds_nb32}.    
    }
    \label{fig:vx_newtonian}
\end{figure}
Non-Newtonian behavior is also observed in the cross-sectional velocity profiles (see Fig.~\ref{fig:vx_newtonian}).
The deviation clearly becomes more pronounced as the pressure difference $\varDelta P$ increases \red{[see Figs.~\ref{fig:vx_newtonian}(a)--\ref{fig:vx_newtonian}(c)]}.
At a large pressure difference, $\varDelta P/\bar{\rho} = 4$ (Fig.~\ref{fig:vx_newtonian}(c)), the velocity gradients of the model polymeric fluid become steeper near the boundaries due to shear thinning.
As a result, the velocity profiles become significantly sharper than those of the Newtonian fluid, particularly in the narrow channel region.
\red{The statistical uncertainty in the time-averaged velocity profiles was estimated by dividing the final sampling interval of $5\times10^6$ MD steps into five equal time blocks.
The standard error was evaluated for each cross-sectional bin from the block-averaged velocities and normalized by the maximum velocity in the corresponding MD cell.
The maximum relative standard errors over all bins and MD cells were $2.1\%$, $1.0\%$, and $1.2\%$ for $\Delta P/\bar{\rho}=1$, $2$, and $4$, respectively.
These small uncertainty levels support the physical significance
of the pronounced deviations from the Newtonian profiles observed
under strong driving.
}

\begin{figure}
    \centering
    \includegraphics[width=0.8\linewidth]{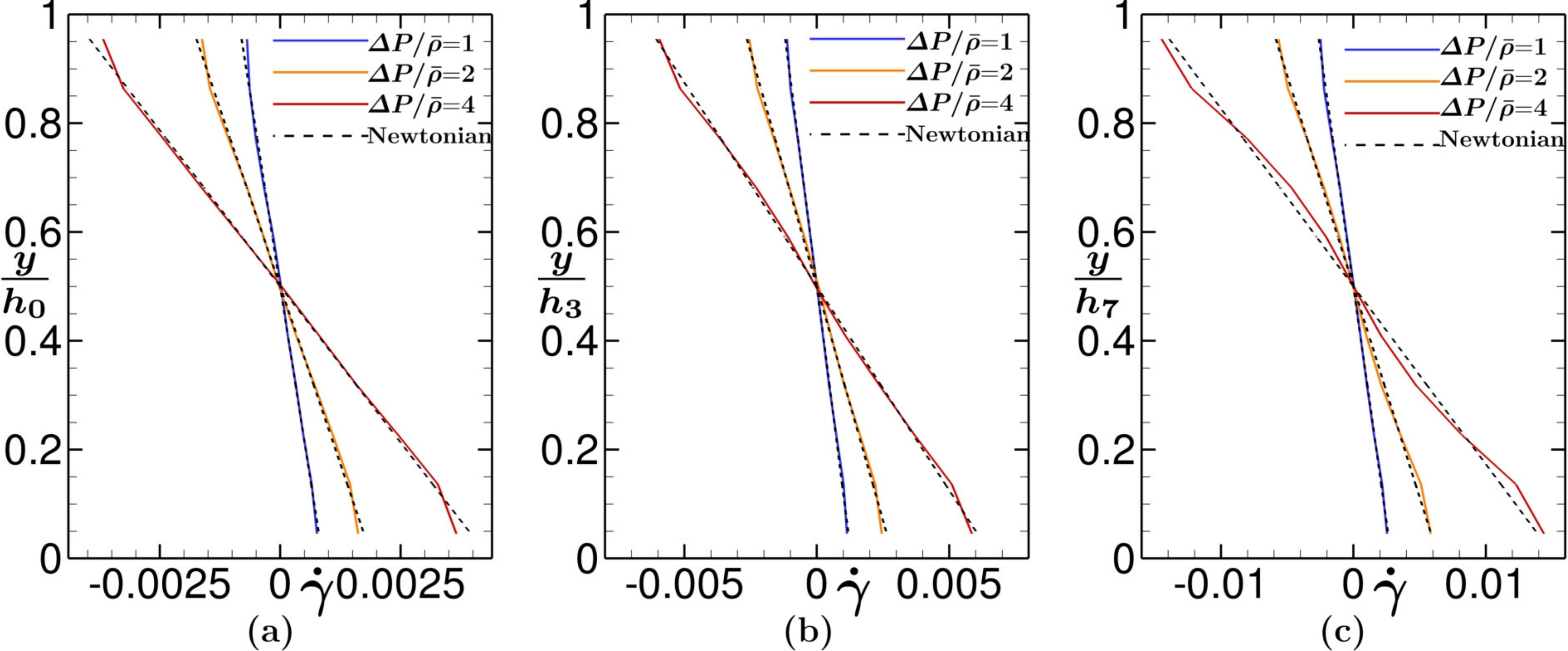}
    \caption{\red{
   Cross-sectional distributions of the local shear rate $\dot{\gamma}=\partial u/\partial y$ at (a) $x=x_0$, (b) $x=x_3$, and (c) $x=x_7$ for pressure-driven flows of the polymeric fluid at $\Delta P/\bar{\rho}=1$, $2$, and $4$.
   The solid lines show the SMD results, and the dashed lines show the corresponding Newtonian-fluid profiles. 
   For the Newtonian reference at $x=x_i$, the shear-rate distribution is evaluated as
   $\dot{\gamma}_{\mathrm{N}}=(F_i/\mu_i)(h_i/2-y)$, using the local driving force $F_i$ and the cell-averaged viscosity $\mu_i$.
   The cross-sectional coordinate is normalized by the local channel height $h_i$. 
    }}
    \label{fig:gamma_newtonian}
\end{figure}
\red{
To examine the deviation from Newtonian behavior more directly, the cross-sectional shear-rate distributions $\dot{\gamma}=\partial u/\partial y$ obtained by the SMD simulations at the representative streamwise positions $x=x_0$, $x_3$, and $x_7$ are compared with those of the Newtonian fluids in Figs.~\ref{fig:gamma_newtonian}(a)--\ref{fig:gamma_newtonian}(c), respectively.
The local shear rate was evaluated using the central difference for the interior bins, as in the definition preceding Eq.~(\ref{eq:calib_viscos}), and one-sided differences for the bins adjacent to the walls.
For a Newtonian fluid under the same local driving force $F_i$ and cell-averaged viscosity $\mu_i$, the shear rate varies linearly across the gap as $\dot{\gamma}_{\mathrm{N}}=(F_i/\mu_i)(h_i/2-y)$.
At the small pressure difference
$\Delta P/\bar{\rho}=1$, the SMD results remain close to the corresponding Newtonian distributions.
As the pressure difference increases, the deviation from the linear Newtonian profiles becomes more pronounced, particularly at $x=x_7$, where the channel is narrowest.
This systematic deviation supports the shear-thinning interpretation of the velocity profiles shown in Fig.~\ref{fig:vx_newtonian}.}

\begin{figure}
    \centering
    \includegraphics[width=0.8\linewidth]{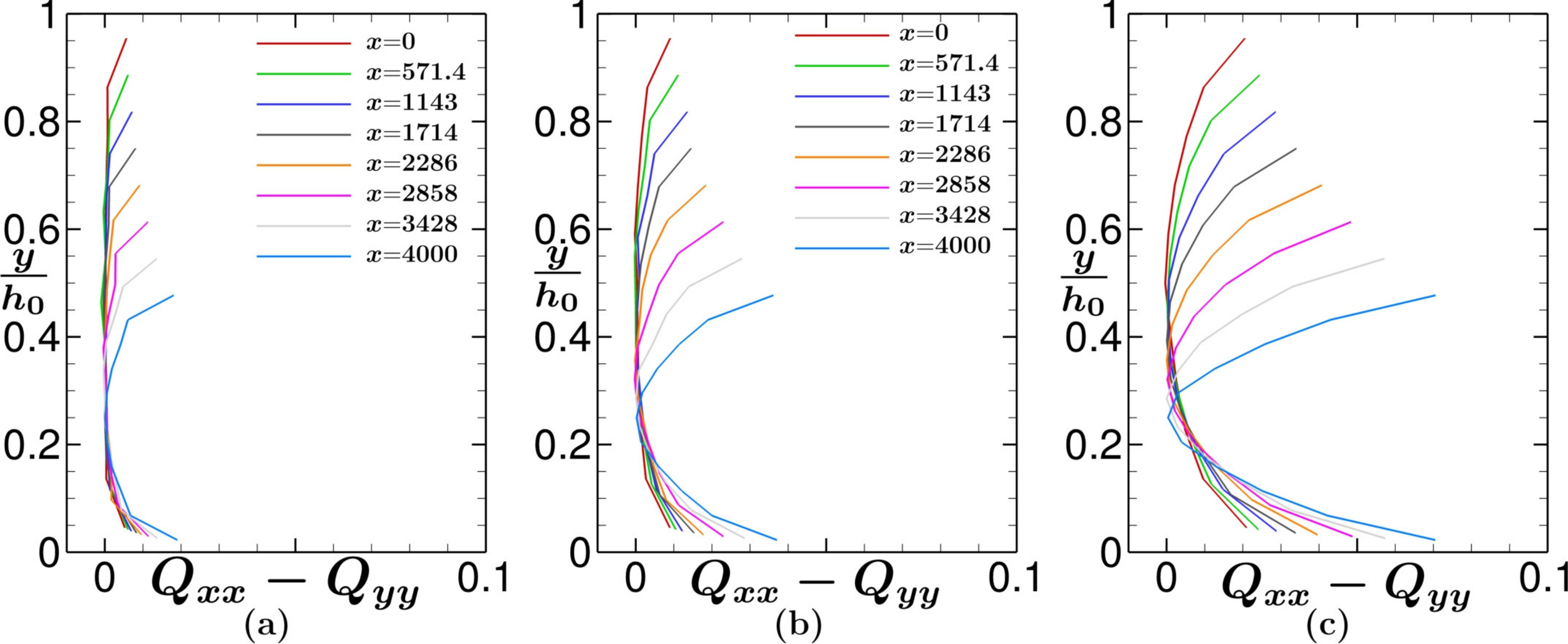}
    \caption{
    Spatial variations in the bond orientation of the model polymeric fluid at different cross sections.
    The left panel [(a)] shows the results for $\varDelta P/\bar\rho = 1$, the middle panel [(b)] for $\varDelta P/\bar\rho = 2$, and the right panel [(c)] for $\varDelta P/\bar\rho = 4$.
    }\label{fig:Q1_profs_nb32}
\end{figure}

The polymer conformation exhibits increasingly pronounced spatial variations as the pressure difference increases.
\red{Figures~\ref{fig:Q1_profs_nb32}(a)--\ref{fig:Q1_profs_nb32}(c) show} the spatial distributions of the anisotropy of the bond orientation tensor, $Q_{xx}-Q_{yy}$, \red{for different pressure differences}.
The bond orientation tensor in the $l$th bin of the MD cell $C_i$, $Q^l_{\alpha\beta,i}$ is calculated as follows:
\begin{equation}\label{eq:qab}
Q_{\alpha\beta,i}^l=\frac{1}{|\varDelta V_i^l|}\sum_{p=1}^{P_i}
\int_{\delta V_i^l}\delta(r^c_{y,p}-r_y)Q_{\alpha\beta,p}d\mathbf{r},
\end{equation}
with
$$
Q_{\alpha\beta,p}=\frac{1}{N_b-1}\sum_{n=1}^{N_b-1}(r_{\alpha,n+1,p}-r_{\alpha,n,p})
(r_{\beta,n+1,p}-r_{\beta,n,p}),
$$
$$
r^c_{\alpha,p}=\frac{1}{N_b}\sum_{n=1}^{N_b}r_{\alpha,n,p},
$$
where $\mathbf{r}_{n,p}$ is the position of the $n$th bead of the $p$th polymer chain, $\mathbf{r}^c_p$ is the center of mass of the $p$th polymer chain, and ${P}_i$ denotes the number of polymer chains contained in the MD cell $C_i$.

At a small pressure difference, $\varDelta P/\bar\rho=1$ (Fig.~\ref{fig:Q1_profs_nb32}(a)), the polymer conformation is nearly uniform, except in the vicinity of the upper and lower boundaries.
In contrast, at a large pressure difference, $\varDelta P/\bar\rho=4$ (Fig.~\ref{fig:Q1_profs_nb32}(c)), the anisotropy of the bond orientation becomes significantly more pronounced throughout the channel, except near the mid-height region. In particular, it increases toward the boundaries.

\red{This near-wall increase in the conformational anisotropy can be understood from the local shear-rate distributions shown in Figs.~\ref{fig:gamma_newtonian}(a)--\ref{fig:gamma_newtonian}(c).
A local Weissenberg number may be defined as $Wi_i(y)=\tau_c|\dot{\gamma}_i(y)|$, which compares the conformational relaxation time $\tau_c$ with the local deformation time $|\dot{\gamma}_i(y)|^{-1}$.
When $Wi_i(y)\gtrsim 1$, the polymer segments are deformed and aligned by the flow faster than they can relax toward an isotropic state. 
Using the estimate $\tau_c\sim 10^3$ discussed in Sec.~IV A, the near-wall shear rates in Fig.~\ref{fig:gamma_newtonian}, which are of the order of $10^{-3}$--$10^{-2}$, correspond approximately to $Wi_i(y)=O(1)$--$O(10)$, whereas the local shear rate approaches zero near the mid-height region.
Therefore, the increase in $|\dot{\gamma}|$ toward the walls and in the narrower downstream region is consistent with the corresponding increase in
$Q_{xx}-Q_{yy}$ shown in Fig.~13.}
\blue{ This observation, that polymers tend to align near the walls as the flow strength increases, is consistent with the typical behavior of polymer melts under shear flow~\cite{cho2017molecular}.}

\begin{figure}
    \centering
    \includegraphics[width=0.8\linewidth]{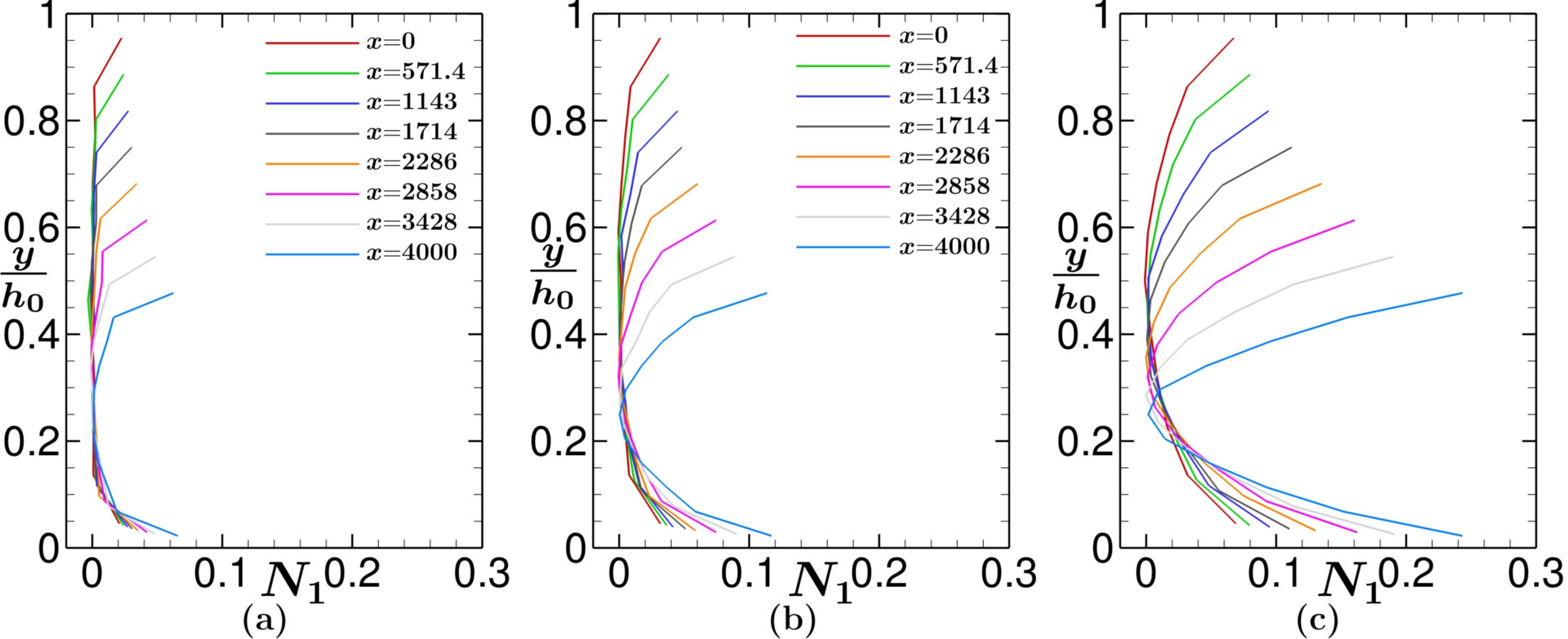}
    \caption{
    Spatial variations in the normal stress difference of the model polymeric fluid at different cross sections.
    The left panel [(a)] shows the results for $\varDelta P/\bar\rho = \red{1}$, the middle panel [(b)] for $\varDelta P/\bar\rho = \red{2}$, and the right panel [(c)] for $\varDelta P/\bar\rho = \red{4}$.
    }
    \label{fig:N1_profs_nb32}
\end{figure}

Consistent with this increased anisotropy of the polymer conformation, the normal stress difference also becomes pronounced in the narrow cross-section at a large difference in pressure (see Fig.~\ref{fig:N1_profs_nb32}).

The dynamics of the polymer conformation at mid-height and in the vicinity of the lower boundary in different MD cells are shown in the movie.

\subsection{Critical analysis}
In the present SMD method, the local MD cells are spatially fixed along the channel.
Thus, the advection of the memory of the polymer conformation, which is transported with the flow velocity, cannot be explicitly traced.
The time required for a polymer to pass through an interval $\varDelta x$ with a characteristic streaming velocity $U_\mathrm{c}$ is estimated as $\varDelta x/U_\mathrm{c}$.
Accordingly, the present SMD method requires the condition
\begin{equation}\label{eq:limitcondition}
    \tau_\mathrm{c}<\varDelta x/U_\mathrm{c},
\end{equation}
where $\tau_\mathrm{c}$ is the characteristic relaxation time of the polymer conformation, under which the advection of the memory of the polymer conformation can be neglected.

The characteristic streaming velocity is estimated as follows:
$$
U_\mathrm{c}=\frac1L\int_0^L \bar{u}(x)dx,
$$
with
$$
\bar u(x)=\frac{1}{h(x)}\int_0^{h(x)} u(x,y)dy.
$$
The resulting values are $U_\mathrm{c}=0.09$, 0.2, and 0.4 for $\varDelta P/\bar\rho=1$, 2, and 4, respectively.
Since the interval $\varDelta x$=571 is fixed in the present simulations, the right-hand side of Eq.~(\ref{eq:limitcondition}) is estimated as $\varDelta x/U_\mathrm{c}$=$6.3\times 10^3$, $2.9\times 10^3$, and $1.4\times 10^3$ for $\varDelta P/\bar\rho=1$, 2, and 4, respectively.
On the other hand, the relaxation time of the KG polymer with $N_\mathrm{b}=32$ is approximately $\tau_\mathrm{c}\sim 1000$~\cite{behbahani2024relaxation,Oda2024}.
Hence, the condition (\ref{eq:limitcondition}) is always satisfied for the present SMD simulations.

However, when the pressure difference becomes larger, the characteristic speed $U_\mathrm{c}$ may no longer satisfy the condition \red{(\ref{eq:limitcondition})}.
Additionally, when the polymer chain becomes large, the characteristic relaxation time $\tau_\mathrm{c}$ may increase such that the condition is violated \red{(\ref{eq:limitcondition})}.
Thus, one must carefully design the simulation configuration, in particular, the mesh interval $\varDelta x$, \red{by balancing the required streamwise resolution, the validity condition  (\ref{eq:limitcondition}), and the computational savings afforded by sparsely distributed MD cells.}

\red{Another limitation of the present formulation concerns its applicability to extremely nanoscale confinements ~\cite{Barisik2014, Zhou2024} as described in the introduction section.
The SMD method proposed here is based on a macroscopic
lubrication description in the streamwise direction, while the cross-sectional molecular dynamics, including the fluid--wall interfaces, is resolved by local MD cells. 
Therefore, the method is intended for thin-layer flows in which the aspect ratio is small and a macroscopic streamwise flux can still be meaningfully defined. 
However, when the gap width becomes comparable to only a few molecular diameters, strong wall-induced inhomogeneities such as molecular layering may dominate the transport. 
In such a regime, the validity of the macroscopic lubrication approximation used in the present SMD formulation is no longer guaranteed.
To examine consistency in that extreme nanoscale confinement, a careful comparison between SMD simulations and direct full MD simulations is required. 
Thus, the present study does not claim applicability to such strongly confined fluids, and extension of SMD to this regime is left as an important future problem.}

\section{Concluding remarks}
In this study, we developed a synchronized molecular dynamics (SMD) framework based on the lubrication approximation for investigating laminar film flows in narrow gaps without relying on constitutive relationships or macroscopic boundary conditions.
By sparsely allocating local molecular dynamics cells along the streamwise direction and synchronizing them through the macroscopic continuity constraint, the proposed method provides a direct multiscale description of confined flows driven by pressure gradients and wall motions.

The validity of the present SMD method was demonstrated through systematic comparisons with the modified Reynolds equation, which incorporates slip boundary conditions, using simple Lennard--Jones fluids as model systems.
For both pressure-driven and wall-driven film flows, the SMD results show excellent agreement with the continuum predictions over a wide range of fluid densities and fluid--wall interaction strengths.
In particular, the method accurately reproduces the spatial distributions of the driving forces and velocity profiles, as well as the slip lengths, which are consistent with kinetic theory expectations in \red{low-density slip flow} regimes.

We also examined the convergent properties of the synchronization scheme and clarified the trade-off between statistical accuracy and communication overhead inherent in the nested parallel implementation.
The results reveal the existence of an optimal number of MD steps per iteration that minimizes fluctuations while maintaining computational efficiency, providing practical guidance for large-scale multiscale simulations.

In addition to the validation for simple Lennard--Jones fluids, the present SMD framework has been further applied to thin-layer flows of polymeric liquids modeled with the Kremer--Grest model.
The results indicate that the SMD method captures non-Newtonian effects, such as shear thinning, which become pronounced at large pressure differences.
In particular, significant deviations from the modified Reynolds equation are observed under strong driving conditions, indicating the breakdown of constant-viscosity assumptions for polymeric fluids.

Moreover, the SMD simulations provide detailed insights into the microscopic origin of these macroscopic deviations.
The spatial variation in polymer conformations, characterized by the anisotropy of the bond orientation tensor, is strongly correlated with the local flow field and shear rate.
These results highlight the ability of the SMD framework to directly resolve the coupling between molecular-scale dynamics and macroscopic transport in non-Newtonian thin-layer flows.

\section*{Supplementary Material}
The dynamics of polymer conformations in different MD cells for pressure-driven flows at $\Delta P = 0.8$, $1.6$, and $3.2$ are provided in Supplementary Movies S1--S3.

\section*{Acknowledgments}
The computations in this work were performed using the supercomputers at Kyoto University and the Center for Cooperative Work on Data Science and Computational Science, University of Hyogo.
This work was supported by JSPS KAKENHI Grant Number JP25K07246.

\appendix
\section{Lubrication approximation}
We consider the two-dimensional incompressible flow of a thin film (i.e., the aspect ratio $\red{\chi} = H/L\ll 1$), as depicted in Fig.~\ref{fig:geom}.
The mass and momentum conservation laws are written as follows:
\begin{subequations}\label{eq_convlaws}
\begin{equation}
\frac{\partial u}{\partial x}+
\frac{\partial v}{\partial y}=0,
\end{equation}
\begin{equation}
    \frac{\partial u}{\partial t}
    +{u}\frac{\partial u}{\partial x}
    +{v}\frac{\partial u}{\partial y}
    =\frac1\rho\left(\frac{\partial \sigma_{xx}}{\partial x}
    +\frac{\partial \sigma_{xy}}{\partial y}\right),
\end{equation}    
\begin{equation}
    \frac{\partial v}{\partial t}
    +{u}\frac{\partial v}{\partial x}
    +{v}\frac{\partial v}{\partial y}
    =\frac{1}{\rho}\left(\frac{\partial \sigma_{xy}}{\partial x}
    +\frac{\partial \sigma_{yy}}{\partial y}\right),
\end{equation}    
\end{subequations}
where $\bm{v}=(u,v)$ is the flow velocity, $t$ is the time, and $\sigma_{\alpha\beta}$ is the stress tensor.
Here, the subscripts $\alpha$ and $\beta$ each represent one of the Cartesian coordinates $\{x,y,z\}$.
We denote the characteristic quantities for velocity, time, pressure, and shear viscosity as $u_0$, $v_0$, $t_0$, $p_0$, and $\mu_0$, respectively, and introduce the following nondimensional quantities:
\begin{gather}
    \hat{x}=\frac{x}{L},\quad 
    \hat{y}=\frac{y}{H},\quad 
    \hat{t}=\frac{t}{t_0},\quad
    \hat{u}=\frac{u}{u_0},\quad
    \hat{v}=\frac{v}{v_0},\quad 
    \hat{\sigma}_{xx,yy}=\frac{\sigma_{xx,yy}}{p_0},\quad
    \hat{\sigma}_{xy}=\frac{\sigma_{xy}}{\frac{\mu_0 u_0}{H}},
\end{gather}
where $\hat{\quad}$ denotes the nondimensional quantity.
Then, Eqs.~(\ref{eq_convlaws}) are written as follows:
\begin{subequations}
\begin{align}
\frac{u_0}{L}\frac{\partial \hat u}{\partial \hat x}+
\frac{v_0}{H}\frac{\partial \hat v}{\partial \hat y}
&=0, \nonumber \\
\frac{\partial \hat u}{\partial \hat x}+
\frac{v_0}{\red{\chi} u_0}\frac{\partial \hat v}{\partial \hat y}
&=0,
\end{align}

\begin{align}
    \frac{u_0}{t_0}\frac{\partial \hat u}{\partial \hat t}
    +\frac{u_0^2}{L}{\hat u}\frac{\partial \hat u}{\partial \hat x}
    +\frac{v_0 u_0}{H}\hat{v}\frac{\partial \hat u}{\partial \hat y}
    =\frac{1}{\hat{\rho}}\left(
    \frac{p_0}{\rho_0 L}\frac{\partial \hat \sigma_{xx}}{\partial \hat x}
    +\frac{\mu_0 u_0}{\rho_0 H^2}\frac{\partial \hat \sigma_{xy}}{\partial \hat y}
    \right), \nonumber\\
    \frac{\rho_0 H^2}{\mu_0 u_0}\left(
     \frac{u_0}{t_0}
     \frac{\partial \hat u}{\partial \hat t}+
    \frac{u_0^2}{L}{\hat u}\frac{\partial \hat u}{\partial \hat x}
    +\frac{v_0 u_0}{H}\hat{v}\frac{\partial \hat u}{\partial \hat y}
    \right)
    =\frac{1}{\hat{\rho}}\left(
    \frac{p_0 H^2}{\mu_0 u_0 L}\frac{\partial \hat \sigma_{xx}}{\partial \hat x}
    +\frac{\partial \hat \sigma_{xy}}{\partial \hat y}
    \right), \nonumber\\
    \frac{\rho_0 H^2}{\mu_0 t_0}\frac{\partial \hat u}{\partial \hat t}
    +\red{\chi} Re \left({\hat u}\frac{\partial \hat u}{\partial \hat x}
    +\frac{v_0}{\red{\chi} u_0}\hat{v}\frac{\partial \hat u}{\partial \hat y}
    \right)
    =\frac{1}{\hat{\rho}}\left(
    \frac{p_0 \red{\chi} H}{\mu_0 u_0}\frac{\partial \hat \sigma_{xx}}{\partial \hat x}
    +\frac{\partial \hat \sigma_{xy}}{\partial \hat y}
    \right), 
\end{align}
\begin{align}
    \frac{v_0}{t_0}\frac{\partial \hat v}{\partial \hat t}
    +\frac{u_0v_0}{L}{\hat u}\frac{\partial \hat v}{\partial \hat x}
    +\frac{v_0^2}{H}{\hat v}\frac{\partial \hat v}{\partial \hat y}
    =\frac{1}{\hat\rho}\left(
    \frac{\mu_0 u_0}{\rho_0 L^2}\frac{\partial \hat \sigma_{xy}}{\partial \hat x}
    +\frac{p_0}{\rho_0 H}\frac{\partial \hat \sigma_{yy}}{\partial \hat y}
    \right),\nonumber \\
    \frac{\rho_0H^2}{\mu_0u_0}\left(
    \frac{v_0}{t_0}\frac{\partial \hat v}{\partial \hat t}
    +\frac{u_0v_0}{L}{\hat u}\frac{\partial \hat v}{\partial \hat x}
    +\frac{v_0^2}{H}{\hat v}\frac{\partial \hat v}{\partial \hat y}
    \right)
    =\frac{1}{\hat\rho}\left(
    \red{\chi}^2\frac{\partial \hat \sigma_{xy}}{\partial \hat x}
    +\frac{p_0 H}{\mu_0 u_0}\frac{\partial \hat \sigma_{yy}}{\partial \hat y}
    \right),\nonumber \\
    \frac{v_0}{u_0}\frac{\rho_0H^2}{\mu_0 t_0}\frac{\partial \hat v}{\partial \hat t}
    +\frac{\red{\chi} Re   v_0}{u_0}{\hat u}\frac{\partial \hat v}{\partial \hat x}
    +\frac{Re v_0^2}{u_0^2}{\hat v}\frac{\partial \hat v}{\partial \hat y}
    =\frac{1}{\hat\rho}\left(
    \red{\chi}^2\frac{\partial \hat \sigma_{xy}}{\partial \hat x}
    +\frac{p_0 H}{\mu_0 u_0}\frac{\partial \hat \sigma_{yy}}{\partial \hat y}
    \right),\nonumber \\
    \frac{v_0}{u_0}\left[
    \frac{\rho_0H^2}{\mu_0 t_0}\frac{\partial \hat v}{\partial \hat t}
    +\red{\chi} Re
    \left(
    {\hat u}\frac{\partial \hat v}{\partial \hat x}
    +\frac{v_0}{\red{\chi} u_0}{\hat v}\frac{\partial \hat v}{\partial \hat y}
    \right)
    \right]
    =\frac{1}{\hat\rho}\left(
    \red{\chi}^2\frac{\partial \hat \sigma_{xy}}{\partial \hat x}
    +\frac{p_0 H}{\mu_0 u_0}\frac{\partial \hat \sigma_{yy}}{\partial \hat y}
    \right),
\end{align}    
\end{subequations}
where $Re$ is the Reynolds number, which is defined as $Re=\frac{\rho_0 u_0 H}{\mu_0}$.
Here, we use $H=\red{\chi} L$.
When we consider $v_0=\red{\chi} u_0$, $t_0=\rho_0H^2/\mu_0$, and $p_0=\mu_0 u_0/(\red{\chi} H)$, we obtain the following equations:
\begin{subequations}
    \begin{equation}
    \frac{\partial \hat u}{\partial \hat x}+
    \frac{\partial \hat v}{\partial \hat y}=0
    \end{equation}
    \begin{equation}
    \frac{\partial \hat u}{\partial \hat t}
    +\red{\chi} Re \left({\hat u}\frac{\partial \hat u}{\partial \hat x}
    +\hat{v}\frac{\partial \hat u}{\partial \hat y}
    \right)
    =\frac{1}{\hat{\rho}}\left(
    \frac{\partial \hat \sigma_{xx}}{\partial \hat x}
    +\frac{\partial \hat \sigma_{xy}}{\partial \hat y}
    \right),     
    \end{equation}
    \begin{equation}
    \red{\chi} \left[
    \frac{\partial \hat v}{\partial \hat t}
    +\red{\chi} Re
    \left(
    {\hat u}\frac{\partial \hat v}{\partial \hat x}
    +{\hat v}\frac{\partial \hat v}{\partial \hat y}
    \right)
    \right]
    =\frac{1}{\hat\rho}\left(
    \red{\chi}^2\frac{\partial \hat \sigma_{xy}}{\partial \hat x}
    +\frac{1}{\red{\chi}}\frac{\partial \hat \sigma_{yy}}{\partial \hat y}
    \right).
    \end{equation}
\end{subequations}
When we ignore the higher-order terms (i.e.,
${\cal O}(\red{\chi}^2)$ and ${\cal O}(\red{\chi} Re)$ terms) and rewrite the equations in the dimensional form, we obtain the lubrication approximation as Eqs.~\eqref{eq_cont}--\eqref{eq_pyy}.

\section{Reynolds equation for slip flows}\label{app:slip}

We consider a Newtonian fluid with a viscosity $\mu$ under a uniform constant force $F$ in the slab geometry.
The steady state of the fluid velocity is described as follows:
\begin{equation}\label{eq_ap2_basic}
    F+\mu\frac{\partial^2 u}{\partial y^2}=0.
\end{equation}
At the upper and lower boundaries, we consider slip flows, which are described as follows:
\begin{align}\label{eq_ap2_boundary}
    &u(y=h)=-\beta\frac{\partial u}{\partial y}+U_\mathrm{u},\\
    &u(y=0)=\beta\frac{\partial u}{\partial y}+U_\mathrm{b},
\end{align}
where $h$ is the thickness of the slab, $U_\mathrm{u}$ and $U_\mathrm{b}$ are the tangential velocities of the upper and lower walls, respectively, and $\beta$ is the slip length.

The local flow velocity in the slab $y=[0,h]$ is described as follows:
\begin{equation}
u(y)=-\frac{F}{\mu}y^2
+\left(\frac{F}{\mu}h+\frac{U_\mathrm{u}-U_\mathrm{b}}{2\beta+h}\right)y
+\beta\left(\frac{F}{\mu}h+\frac{U_\mathrm{u}-U_\mathrm{b}}{2\beta+h}\right)+U_\mathrm{b}.
\end{equation}
By integrating the above equation, we obtain
\begin{equation}
M=\int_0^hu(y)dy=\frac{Fh^2}{12\mu}(h+6\beta)+\frac{h}{2}(U_\mathrm{u}+U_\mathrm{b}).
\end{equation}

Using the above equation as the local flux in Eq.~(\ref{eq_cont2}), we can obtain the modified Reynolds equation for slip boundary flows, which determines the spatial distribution of $F(x)$.
For example, for the case where the boundary walls move only in their tangential directions (i.e., $V_\mathrm{u}/U_\mathrm{u}=h'(x)$, and $V_\mathrm{b}=0$), the right-hand side of Eq.~(\ref{eq_cont2}) vanishes.
Thus, the distribution of $F$ is described as follows:
\begin{equation}\label{eq:reynolds_slip}
F(x)=\frac{C}{h(x)^2(h(x)+\beta)}-\frac{12\mu\bar u}{h(x)(h(x)+6\beta)},
\end{equation}
where $C$ is an undetermined constant and $\bar u=\frac12(U_\mathrm{u}+U_\mathrm{b})$.

The constant $C$ can be calculated from the difference in pressure $\varDelta P$ between the inlet and outlet of the channel, i.e.,
$$
\int_0^L F(x) dx=-\varDelta P.
$$
For example, for the wedge-shaped channel with $h'(x)=(h_I-h_0)/L$ depicted in Fig.~\ref{fig:geom}, we have
\begin{equation}\label{eq:C_reynolds_slip}
C=6\beta\left\{
\frac{1}{h_0}-\frac{1}{h_I}
-\frac{1}{6\beta}\log\left(\frac{1+\frac{6\beta}{h_0}}{1+\frac{6\beta}{h_I}}\right)
\right\}^{-1}
\left\{
\frac{h_I-h_0}{L}\varDelta P+\frac{2\mu\bar u}{\beta}
\log\left(\frac{1+\frac{6\beta}{h_0}}{1+\frac{6\beta}{h_I}}\right)
\right\}.
\end{equation}
\bibliography{SMD_lubric.bib}

\end{document}